\documentclass[zpreprint,zbstdefault]{zeus_paper}
\usepackage[english]{babel}
\chardef\usc=95
\chardef\til=126
\catcode`\@=11 
\DeclareRobustCommand\xdotspace{\futurelet\@let@token\@xdotspace}
\def\@xdotspace{%
  \ifx\@let@token.\else
  \ifx\@let@token\bgroup.\else
  \ifx\@let@token\egroup.\else
  \ifx\@let@token\/.\else
  \ifx\@let@token\ .\else
  \ifx\@let@token~.\else
  \ifx\@let@token!.\else
  \ifx\@let@token,.\else
  \ifx\@let@token:.\else
  \ifx\@let@token;.\else
  \ifx\@let@token?.\else
  \ifx\@let@token/.\else
  \ifx\@let@token'.\else
  \ifx\@let@token).\else
  \ifx\@let@token-.\else
  \ifx\@let@token\@xobeysp.\else
  \ifx\@let@token\space.\else
  \ifx\@let@token\@sptoken.\else
   .\space
   \fi\fi\fi\fi\fi\fi\fi\fi\fi\fi\fi\fi\fi\fi\fi\fi\fi\fi}
\catcode`\@=12 

\newcommand{\stru}[2]{%
   \relax\ifmmode\hbox{\vrule height#1 depth#2 width0pt}%
   \else\vrule height#1 depth#2 width0pt\fi}

\newcommand{\Ronum}[1]{\uppercase\expandafter{\romannumeral#1}}
\newcommand{\ronum}[1]{\expandafter{\romannumeral#1}}
\DeclareRobustCommand{\LaTeXZ}{%
  \LaTeX\kern-.05em4\kern-.1em
  {\raisebox{-0.2ex}{$\scriptstyle\text{ZEUS}$}}\xspace}

\DeclareMathAlphabet{\mathbf}{OT1}{cmr}{bx}{sl}
\newcommand{\eVdist}{\kern-0.06667em}

\newcommand{\gev}{{\,\text{Ge}\eVdist\text{V\/}}}

\newcommand{\nb}{\,\text{nb}}

\newcommand{\cm}{\,\text{cm}}

\newcommand{\Tesla}{\,\text{T}}

\newcommand{\slashfrac}[2]{%
  \raisebox{0.5ex}{\ensuremath #1}\kern-0.12em/\kern-0.08em
  \raisebox{-.8ex}{\ensuremath #2}}

\newcommand{\sqr}[3]{%
    {\vcenter{\hrule height.#3ex\hbox{\vrule width.#2ex height#1ex
     \kern#1ex\vrule width.#3ex}\hrule height.#2ex}}}

\catcode`\@=11 
\newcommand{\parenbar}{\mathpalette\p@renb@r}
\def\p@renb@r#1#2{\vbox{%
  \ifx#1\scriptscriptstyle \dimen@.7em\dimen@ii.2em\else
  \ifx#1\scriptstyle \dimen@.8em\dimen@ii.25em\else
  \dimen@1em\dimen@ii.4em\fi\fi \offinterlineskip
  \ialign{\hfill##\hfill\cr
    \vbox{\hrule width\dimen@ii}\cr
    \noalign{\vskip-.3ex}%
    \hbox to\dimen@{$\mathchar300\hfil\mathchar301$}\cr
    \noalign{\vskip-.3ex}%
    $#1#2$\cr}}}
\catcode`\@=12 

\newcommand{\IP}{{\rm I$\kern-0.01667em$P}\xspace}

\mathchardef\qsm=63
\mathchardef\pls=43
\mathchardef\mns=512
\mathchardef\plm=518
\mathchardef\eql=61
\mathchardef\smallleft=300
\mathchardef\smallright=301
\mathchardef\les=316
\mathchardef\gre=318
\mathchardef\leq=532
\mathchardef\grq=533

\catcode`\@=11 
\newcounter{pict@width}
\newcounter{pict@height}
\newlength{\pict@scale}
\newcommand{\psfigadd}[4]{%
\setcounter{pict@width}{1*\ratio{#2+\pict@scale/2}{\pict@scale}}
\setcounter{pict@height}{1*\ratio{#3+\pict@scale/2}{\pict@scale}}
\setlength{\unitlength}{\pict@scale}
\hbox to #2{\hspace{-\fill}\begin{picture}(\thepict@width,\thepict@height)
\put(0,0){\psfig{figure=#1,width=#2,height=#3,clip=}}
\SetScale{0.283466457}
\SetWidth{1.763889}
{#4}
\end{picture}}
}
\newcounter{pict@widthfst}
\newcounter{pict@widthscd}
\newcounter{pict@widthtot}
\newcommand{\psfigaddtwo}[7]{%
\setcounter{pict@widthfst}{1*\ratio{#2+\pict@scale/2}{\pict@scale}}
\setcounter{pict@widthscd}{1*\ratio{#2+#4+\pict@scale/2}{\pict@scale}}
\setcounter{pict@widthtot}{1*\ratio{#2+#4+#6+\pict@scale/2}{\pict@scale}}
\setcounter{pict@height}{1*\ratio{#3+\pict@scale/2}{\pict@scale}}
\setlength{\unitlength}{\pict@scale}
\hbox{\hspace{-\fill}\begin{picture}(\thepict@widthtot,\thepict@height)
\put(0,0){\psfig{figure=#1,width=#2,height=#3,clip=}}
\put(\thepict@widthscd,0){\psfig{figure=#5,width=#6,height=#3,clip=}}
\SetScale{0.283466457}
\SetWidth{1.763889}
{#7}
\end{picture}}
}
\newcommand{\psfigror}[4]{%
\setcounter{pict@width}{1*\ratio{#2+\pict@scale/2}{\pict@scale}}
\setcounter{pict@height}{1*\ratio{#3+\pict@scale/2}{\pict@scale}}
\setlength{\unitlength}{\pict@scale}
\hbox{\begin{picture}(\thepict@width,\thepict@height)
\put(0,\thepict@height){\psfig{figure=#1,width=#3,height=#2,clip=,angle=270}}
\SetScale{0.283466457}
\SetWidth{1.763889}
{#4}
\end{picture}}
}
\newcommand{\psfigrol}[4]{%
\setcounter{pict@width}{1*\ratio{#2+\pict@scale/2}{\pict@scale}}
\setcounter{pict@height}{1*\ratio{#3+\pict@scale/2}{\pict@scale}}
\setlength{\unitlength}{\pict@scale}
\hbox{\begin{picture}(\thepict@width,\thepict@height)
\put(0,0){\psfig{figure=#1,width=#3,height=#2,clip=,angle=90}}
\SetScale{0.283466457}
\SetWidth{1.763889}
{#4}
\end{picture}}
}
\catcode`\@=12 
\newlength\listtextwidth

\catcode`\@=11 
\newlength{\@tabfninsert}
\newlength{\@tabfnwidth}
\newcommand{\tabfootnote}[2]{%
  \setlength{\@tabfninsert}{0.8em}
  \setlength{\@tabfnwidth}{\textwidth}
  \addtolength{\@tabfnwidth}{-\@tabfninsert}
  \addtolength{\@tabfnwidth}{-0.4em}
  \noindent\makebox[\@tabfninsert][r]{\footnotesize$^{#1}$\hfil}\hfill%
  \parbox[t]{\@tabfnwidth}{\footnotesize #2\hfill}}
\catcode`\@=12 
\newcommand{\etgap}{$E_{\rm{T}}^{\rm{GAP}}$}
\newcommand{\etgapmath}{E_{\rm{T}}^{\rm{GAP}}}

\newcommand{\etcutmath}{E_{\rm{T}}^{\rm{CUT}}}

\newcommand{\ptmin}{p_{\rm T}^{\rm min}} 
\newcommand{\ptmina}{p_{\rm T}^{\rm min 1}} 
\newcommand{\ptminb}{p_{\rm T}^{\rm min 2}} 
\newcommand{\xgom}{x_{\gamma}^{\rm OBS}}

\begin{document}
\prepnum{DESY--06--215}

\title{Photoproduction of events with rapidity gaps between jets at HERA}                
                    
\author{ZEUS Collaboration}
\draftversion{After Reading 0.1}
\date{\today}

\confname{ZEUS Paper Draft}
\confplacedate{}
\confsession{}
\confabsnum{}

\abstract{
The photoproduction of dijet events, where the two
jets with the highest transverse energy are separated by a large gap
in pseudorapidity, have been studied with the ZEUS detector using an
integrated luminosity of 39 pb$^{-1}$.  Rapidity-gap events are defined
in terms of the energy flow between the jets, such that the total summed
transverse energy in this region is less than some value $E_{\rm T}^{\rm CUT}$.
The data show a clear excess over the predictions 
of standard photoproduction models. This is interpreted as evidence for 
a strongly interacting exchange of a color-singlet object. Monte Carlo 
models which 
include such a color-singlet exchange are able to describe the data.
}

\makezeustitle

\pagenumbering{Roman}                                                                              
                                                   %
\begin{center}                                                                                     
{                      \Large  The ZEUS Collaboration              }                               
\end{center}                                                                                       
  S.~Chekanov$^{   1}$,                                                                            
  M.~Derrick,                                                                                      
  S.~Magill,                                                                                       
  S.~Miglioranzi$^{   2}$,                                                                         
  B.~Musgrave,                                                                                     
  D.~Nicholass$^{   2}$,                                                                           
  \mbox{J.~Repond},                                                                                
  R.~Yoshida\\                                                                                     
 {\it Argonne National Laboratory, Argonne, Illinois 60439-4815}, USA~$^{n}$                       
\par \filbreak                                                                                     
  M.C.K.~Mattingly \\                                                                              
 {\it Andrews University, Berrien Springs, Michigan 49104-0380}, USA                               
\par \filbreak                                                                                     
  M.~Jechow, N.~Pavel~$^{\dagger}$, A.G.~Yag\"ues Molina \\                                        
  {\it Institut f\"ur Physik der Humboldt-Universit\"at zu Berlin,                                 
           Berlin, Germany}
% ...
\newpage                                                                                           
$^{\    1}$ supported by DESY, Germany \\                                                          
$^{\    2}$ also affiliated with University College London, UK \\                                  
% ...
$^{\dagger}$ deceased \\                                                                           
%
\newpage   
                                                           %
                                                           %
\begin{tabular}[h]{rp{14cm}}                                                                       
$^{a}$ &  supported by the Natural Sciences and Engineering Research Council of Canada (NSERC) \\  
$^{b}$ &  supported by the German Federal Ministry for Education and Research (BMBF), under        
          contract numbers HZ1GUA 2, HZ1GUB 0, HZ1PDA 5, HZ1VFA 5\\                                
$^{c}$ &  supported in part by the MINERVA Gesellschaft f\"ur Forschung GmbH, the Israel Science   
          Foundation (grant no. 293/02-11.2) and the U.S.-Israel Binational Science Foundation \\  
$^{d}$ &  supported by the German-Israeli Foundation and the Israel Science Foundation\\           
$^{e}$ &  supported by the Italian National Institute for Nuclear Physics (INFN) \\                
$^{f}$ &  supported by the Japanese Ministry of Education, Culture, Sports, Science and Technology 
          (MEXT) and its grants for Scientific Research\\                                          
$^{g}$ &  supported by the Korean Ministry of Education and Korea Science and Engineering          
          Foundation\\                                                                             
$^{h}$ &  supported by the Netherlands Foundation for Research on Matter (FOM)\\                   
$^{i}$ &  supported by the Polish State Committee for Scientific Research, grant no.               
          620/E-77/SPB/DESY/P-03/DZ 117/2003-2005 and grant no. 1P03B07427/2004-2006\\             
$^{j}$ &  partially supported by the German Federal Ministry for Education and Research (BMBF)\\   
$^{k}$ &  supported by RF Presidential grant N 1685.2003.2 for the leading scientific schools and  
          by the Russian Ministry of Education and Science through its grant for Scientific        
          Research on High Energy Physics\\                                                        
$^{l}$ &  supported by the Spanish Ministry of Education and Science through funds provided by     
          CICYT\\                                                                                  
$^{m}$ &  supported by the Particle Physics and Astronomy Research Council, UK\\                   
$^{n}$ &  supported by the US Department of Energy\\                                               
$^{o}$ &  supported by the US National Science Foundation. Any opinion,                            
findings and conclusions or recommendations expressed in this material                             
are those of the authors and do not necessarily reflect the views of the                           
National Science Foundation.\\                                                                     
$^{p}$ &  supported by the Polish Ministry of Science and Higher Education\\                       
$^{q}$ &  supported by FNRS and its associated funds (IISN and FRIA) and by an Inter-University    
          Attraction Poles Programme subsidised by the Belgian Federal Science Policy Office\\     
$^{r}$ &  supported by the Malaysian Ministry of Science, Technology and                           
Innovation/Akademi Sains Malaysia grant SAGA 66-02-03-0048\\                                       
\end{tabular}                                                                                      
                                                           %
                                                           %

\pagenumbering{arabic} 
\pagestyle{plain}
\section{Introduction}
\label{sec-int}

The production of events in hadronic collisions with two high transverse energy
jets in the final state separated by a large rapidity interval provides an ideal 
environment to study the interplay between soft (non-perturbative)   
and hard (perturbative) QCD.  

The dominant mechanism for the production of jets with high transverse energy 
in hadronic collisions is a hard interaction between partons in the incoming 
hadrons via a quark or gluon propagator. The exchange of color quantum numbers
generally gives rise to jets in the final state that are color connected to 
each other and to the remnants of the incoming hadrons. 
This leads to energy flow populating 
the pseudorapidity\footnote{The pseudorapidity $\eta=-\ln[\tan(\theta/2)]$, where
$\theta$ is a polar angle.} region both between the jets and the hadronic remnants, 
and between the 
jets themselves. The fraction of events with little or no hadronic activity between the 
jets is expected to be exponentially suppressed as the rapidity interval between the jets increases. A 
non-exponentially suppressed fraction of such events would therefore be 
a signature of the exchange of a color-singlet (CS) object.

The high transverse energy of the jets provides a perturbative hard scale 
at each end of the CS exchange, so that the cross section should 
be calculable in 
perturbative QCD \cite{epj:C1:285}.

Previous studies of jets with rapidity gaps have been made in $p\bar{p}$ collisions at the Tevatron 
\cite{prl:72:2332,*prl:76:734,*pl:B440:189,prl:74:855,*prl:80:1156,*prl:81:5278} 
and in photoproduction at HERA~\cite{pl:b369:55,epj:c24:517}, where a quasi-real
photon from the incoming positron interacts with the proton.  
Comparison with different Monte Carlo (MC) models  
suggested that some contribution of a strong CS 
exchange is required to describe the data, although the uncertainty on the contribution 
from standard QCD processes was large.

In the analysis presented in this paper, photoproduction of dijet events with 
a large rapidity gap between 
jets
is used to investigate the dynamics of color singlet exchange. 
The results are based on a larger data sample, than in the 
previous publications~\cite{pl:b369:55,epj:c24:517}.
The MC models were tuned to better describe the data sample at the detector level.
The CS contribution
is studied and compared to MC models as a function of several kinematic variables and to
a recent QCD-resummed calculation~\cite{newAppleby,pl:b628:49,JHEP:309:56}.

\section{Experimental set-up}
\label{sec-det}
The results presented in this paper correspond to 38.6 $\pm$ 1.6 pb$^{-1}$ of data 
taken with the ZEUS detector
during the 1996-1997 HERA running period. Positrons of 27.5 
$\gev$ collided with protons of 820 $\gev$, giving a center-of-mass energy of $\sqrt{s} = 300 \gev$.

A detailed description of the ZEUS detector
can be found elsewhere~\cite{pl:b293:465,zeus:1993:bluebook}. A brief
outline of the components that are most relevant for this analysis is
given below.

Charged particles are measured in the central tracking detector
(CTD)~\cite{nim:a279:290,*npps:b32:181,*nim:a338:254}, which operates in a
magnetic field of $1.43\Tesla$ provided by a thin super-conducting
solenoid. The CTD consists of 72~cylindrical drift chamber
layers, organized in nine super-layers covering the
polar-angle\footnote{The 
ZEUS coordinate system is a right-handed Cartesian system, with the $Z$ axis 
pointing in the proton beam direction, referred to as the 
``forward direction'', and the $X$ axis pointing left towards the center of 
HERA. The coordinate origin is at the nominal interaction point.} 
region \mbox{$15^\circ<\theta<164^\circ$}. The transverse momentum
resolution for full-length tracks can be parameterized as
$\sigma(p_T)/p_T=0.0058p_T\oplus0.0065\oplus0.0014/p_T$, with $p_T$ in
$\gev$. The tracking system was used to measure the interaction vertex
with a typical resolution along (transverse to) the beam direction of
0.4~(0.1)~cm and also to cross-check the energy scale of the calorimeter.

The high-resolution uranium-scintillator calorimeter
(CAL)~\cite{nim:a309:77,*nim:a309:101,*nim:a321:356,*nim:a336:23} covers
$99.7\%$ of the total solid angle and consists of three parts: the
forward (FCAL), the barrel (BCAL) and the rear (RCAL) calorimeters. Each
part is subdivided transversely into towers and longitudinally into one
electromagnetic section and either one (in RCAL) or two (in BCAL and
FCAL) hadronic sections. The smallest subdivision of the calorimeter
is called a cell. Under test-beam conditions, the CAL single-particle
relative energy resolutions were $\sigma(E)/E=0.18/\sqrt{E}$ for
electrons and $\sigma(E)/E=0.35/\sqrt{E}$ for hadrons, with $E$ in $\gev$.

The luminosity was measured from the rate of the bremsstrahlung process
$ep\rightarrow e\gamma p$. The resulting small angle energetic photons
were measured by the luminosity
monitor~\cite{desy-92-066,*zfp:c63:391,*acpp:b32:2025}, a
lead-scintillator calorimeter placed in the HERA tunnel at $Z=-107$ m.

\section{Kinematics and event selection}

\label{sec-dat}
A three-level trigger system was used to select events online 
\cite{zeus:1993:bluebook,proc:chep:1992:222}.
In the third-level trigger, jets were required to have a transverse energy of
$E_{\rm T}^{\rm jet} > 4 \gev$ 
and a pseudorapidity of $\eta^{\rm jet} < 2.5$ in the laboratory frame.

The $\gamma p$ center-of-mass energy, $W$, and the inelasticity, $y=W^2/s$,
were reconstructed using the Jacquet-Blondel (JB) \cite{proc:epfacility:1979:391} method.
The hadronic system was reconstructed using Energy Flow Objects (EFOs), 
which were formed by combining information 
from energy clusters reconstructed in the CAL and charged tracks reconstructed in the CTD.
The electron ($e$)~\cite{proc:hera:1991:23} reconstruction method was 
also used, in order to remove deep inelastic scattering (DIS) events.

The photoproduction sample was selected by applying the following offline cuts:
\begin{itemize}
\item the longitudinal position of the reconstructed vertex 
was required to be in the range $-40 \cm < Z_{VTX} < 40 \cm$;
\item events with a scattered positron in the CAL having $y_e < 0.85$ and $E^{\prime}_e > 5 \gev$, 
where $E^{\prime}_e$ 
is the energy of the scattered positron, were rejected. This cut reduced contamination 
from neutral current DIS events, since the  
efficiency for the detection of the scattered positron in this region approached 100\%;
\item events were required to  have $0.2 < y_{\rm JB} < 0.75$.  The upper cut on
$y_{\rm JB}$ further reduced contamination
from the neutral current DIS events
that were not removed by the cut on $y_e$ and the lower cut removed beam-gas events;
\item in order
to reduce contributions from charged current events and cosmic-ray showers,
events were required to have a relative transverse momentum 
$P_{\rm T}^{\rm miss} / \sqrt{E_{\rm T}} < 2.0 \gev^{1/2}$, where $P_{\rm T}^{\rm miss}$ and
$E_{\rm T}$ are the total event missing momentum and transverse energy, respectively.
\end{itemize}

The cuts on $y_e$ and $y_{\rm JB}$ reduced the contribution of DIS events to 
less than 0.5\%, confined the phase-space region of the analysis to
$0.2 < y < 0.75$ and
restricted the photon virtuality to a range of $Q^2 < 1 \gev^2$ with a median value of $Q^2 \sim 10^{-3} \gev^2$ \cite{Chekanov:2001bw}.

Jets were reconstructed from the EFOs using the $k_T$ algorithm \cite{np:b406:187} in the longitudinally 
invariant inclusive mode \cite{pr:d48:3160}, which implies that any particle
is included in one of the jets, and ordered 
in $E_{\rm T}^{\rm jet}$, such that jet1 had the 
highest $E_{\rm T}^{\rm jet}$. Events in which 
jets satisfied the following criteria were then selected:
\begin{itemize}
\item jet transverse energy corresponding to 
$E^{\rm jet1}_{\rm T} \ge 6 \gev$ and  $E^{\rm jet2}_{\rm T} \ge 5 \gev$
at the hadron level, after taking in account energy loss in inactive material and other detector effects;
\item $-2.4 < \eta^{\rm jet1,2} <  2.4$, where $\eta^{\rm jet1}$ and 
$\eta^{\rm jet2}$ are the pseudorapidities of the corresponding jets,
to ensure that the jets were well reconstructed in the detector;
\item $2.5 < \Delta \eta < 4$, 
where $\Delta\eta \equiv |\eta^{\rm jet1} - \eta^{\rm jet2}|$ is the absolute difference in 
pseudorapidity between the jets;
\item $\frac{1}{2}|\left(\eta^{\rm jet1} + \eta^{\rm jet2}\right)| <  0.75$, where this condition,
 together with the previous one, constrained the jets to lie within the kinematic region where the detector
and event simulation are well understood.
\end{itemize}

The transverse energy in the gap, \etgap, was calculated by summing up the transverse energy of all jets, without any cut
on $E_{\rm T}^{\rm jet}$, lying in the pseudorapidity region between the 
two highest-$E_{\rm T}^{\rm jet}$ jets satisfying the above requirements \cite{Oderda:1998en}.
Gap events were defined as those in which $\etgapmath$ was less than an $\etcutmath$ value. The $\etcutmath$ values
used in this analysis were $\etcutmath = 0.5, 1.0, 1.5,$ and $2.0 \gev$.
The gap fraction, $f$, was defined as the ratio of the cross section for gap events to the cross section
for inclusive events, which pass all of the above cuts but have no restriction on the $\etgapmath$ value.

In addition, the fraction of the photon momentum  participating
in the hard interaction was calculated as 
$\xgom = (E_{\rm T}^{\rm jet1} e^{-\eta^{\rm jet1}}+E_{\rm T}^{\rm jet2}
e^{-\eta^{\rm jet2}})/ 2yE_e$, where $E_e$ is the energy of the positron beam.  


\section{QCD models and event simulation}

\subsection{Monte Carlo models}

The {\sc Pythia 6.1} \cite{cpc:82:74} and {\sc Herwig 6.1} \cite{cpc:67:465} MC generators
were used to correct the data to the hadron level and for model comparisons. 
Both MCs are based on the leading order (LO) ($2 \rightarrow 2$) matrix elements together with
a parton-shower simulation of additional QCD radiation and hadronisation models. 
The detector
simulation was performed with the {\sc Geant}~3.13 program~\cite{tech:cern-dd-ee-84-1}.

In photoproduction interactions at LO, the photon can either
participate directly in the hard sub-process
(direct photoproduction) or first fluctuate into a hadronic state which then interacts via a partonic constituent carrying some 
fraction, $x_{\gamma}$, of the photon momentum (resolved photoproduction).  
At leading order, therefore, CS exchange between jets may take place only in resolved photoproduction.
For this analysis the direct, resolved, and CS
exchange MC samples were generated separately.

The simulation of multi-parton interactions (MPI) was included in
{\sc Pythia} using the so-called ``simple mode''~\cite{cpc:82:74}
and in {\sc Herwig} by interfacing to 
the {\sc Jimmy} library \cite{zfp:c72:637}.
The minimum transverse momenta, $\ptmin$, of the outgoing partons in the 
hard interaction and partons participating in 
MPI are separately adjustable in {\sc Pythia}, while in {\sc Herwig}
the same parameter was used to adjust both momenta.
The starting parameters for the tuning 
were taken from global fits of {\sc JetWeb} \cite{Butterworth:2002ts}.
The $\ptmin$ was tuned ~\cite{thesis:pryan:2006,thesis:gwenlan:2003} 
for both MC programs by comparing to the data sample 
after the kinematic 
cuts were applied (see Section \ref{sec-dat}).  
The best fit resulted in $\ptmin$ values of
$\ptmina = 1.9 \gev$ and $\ptminb = 1.7 \gev$ for {\sc Pythia} and 
$\ptmin = 2.7 \gev$ for {\sc Herwig}.
For both MC models the CTEQ5L parametrisation~\cite{epj:c12:375} for the proton and the
SaS-G 2D parametrisation~\cite{zfp:c68:607} for the photon PDFs were used. 
Hadronisation in {\sc Herwig} is simulated using the cluster model~\cite{np:b310:461} while {\sc Pythia} uses 
the Lund string model~\cite{prep:97:31}.

The CS exchange is implemented in 
{\sc Herwig} using the LLA BFKL model by Mueller and Tang \cite{pl:b284:123}.
The hard-Pomeron intercept, $1+\omega_0$, 
is related to the strong coupling, $\alpha_s$, used in
the BFKL parton evolution by $\omega_0 = \frac {\alpha_s C_A} {\pi} \left[4 \ln\left(2\right) \right]$.
In this analysis, the default value of $\omega_0 = 0.3$ was used.  

{\sc Pythia} does not contain a simulation of strongly interacting 
CS exchange in hard interactions.  
However, a similar topology 
can be simulated by high-$t$ photon exchange for quark-quark scattering in LO resolved processes.  
Such an exchange 
is not expected to represent
the mechanism of strongly-interacting CS exchange and
is only used to compare the data to an alternative CS model.

\subsection{Resummed calculation}

The gap definition in terms of the energy flow between jets, being infrared safe,
allows pQCD calculations to be applied. These calculations involve the
resummation of large 
logarithms of $E_{\rm T}^{\rm GAP}/E_{\rm T}^{\rm jet}$.
There are several sources of these large logarithms. 
The primary leading logarithms arise
from soft gluon emission directly into the gap, whereas 
secondary (non-global) leading logarithms
are due to emission into the gap from a coherent ensemble of gluons 
outside the gap region~\cite{Kidonakis:1998bk,*np:b308:833,*Sotiropoulos:1993rd,*Contopanagos:1996nh,
JHEP:309:56,pl:b628:49,hep-ph-0604094,jhep:12:063}.

The calculation~\cite{newAppleby} used in this paper provides a prediction of the gap fraction 
with primary emission resummed to all orders and a
correction applied
for the effect of the clustering algorithm, and the non-global 
logarithms correct in the limit of large number of
colors. 
The theoretical uncertainty in this calculation is estimated from varying the 
renormalisation scale between $E_{\rm T} /2$ and $2E_{\rm T}$, where
$E_{\rm T}$ is the transverse energy of the hardest jet.

\section{Data correction and systematic uncertainties}
The data were corrected to the hadron level, bin-by-bin, using correction
factors obtained from a combination of direct, 
resolved, and CS MC samples as described in detail elsewhere 
~\cite{thesis:pryan:2006,thesis:gwenlan:2003}.

The admixture of direct and resolved MC used in the unfolding was determined by the best fit to 
the $\xgom$ data distribution.
The combination of direct and resolved MC formed the non-color-singlet (NCS) sample.

The relative amounts of NCS and CS MC used in the unfolding were determined by the best
fit to the total energy in the gap for events in which $\etgapmath < 1.5 \gev$, after the normalisation of the
NCS sample was fixed using data at $\etgapmath > 1.5 \gev$.  
Fitting to the total number of jets in the gap for events in which $\etgapmath < 1.5 \gev$ and to
$d\sigma/ d\etgapmath$ gave similar results.

To correct the data the average correction factor of {\sc Pythia} 
and {\sc Herwig} was used.
One half of the difference between those two models predictions, about 5\%,
 was assigned to the systematic uncertainties.

A detailed study of the sources contributing to the systematic uncertainties of the measurements 
was performed using {\sc Herwig}.
The analysis cuts were varied by their respective resolutions estimated 
using Monte Carlo.

The variation of the cuts on $\etgapmath$ and $E_{\rm T}$ 
caused the largest contributions to the systematic uncertainty.  
Depending upon the variable measured, their contribution ranged from a few to approximately $30\%$ 
in regions where the statistical significance was low.

The amount of CS exchange MC used in the unfolding was varied by $\pm 25\%$, 
resulting in a variation in the cross section at the one percent level.
All the above systematics were added in quadrature in order to calculate the total systematic
uncertainty.

The calorimeter energy scale was varied by $\pm 3\%$. 
This uncertainty was not combined with the other systematics, but 
instead shown separately as a shaded band in the figures.

\section{Results}
\label{results}

\begin{figure}
\centerline{\epsfig{file=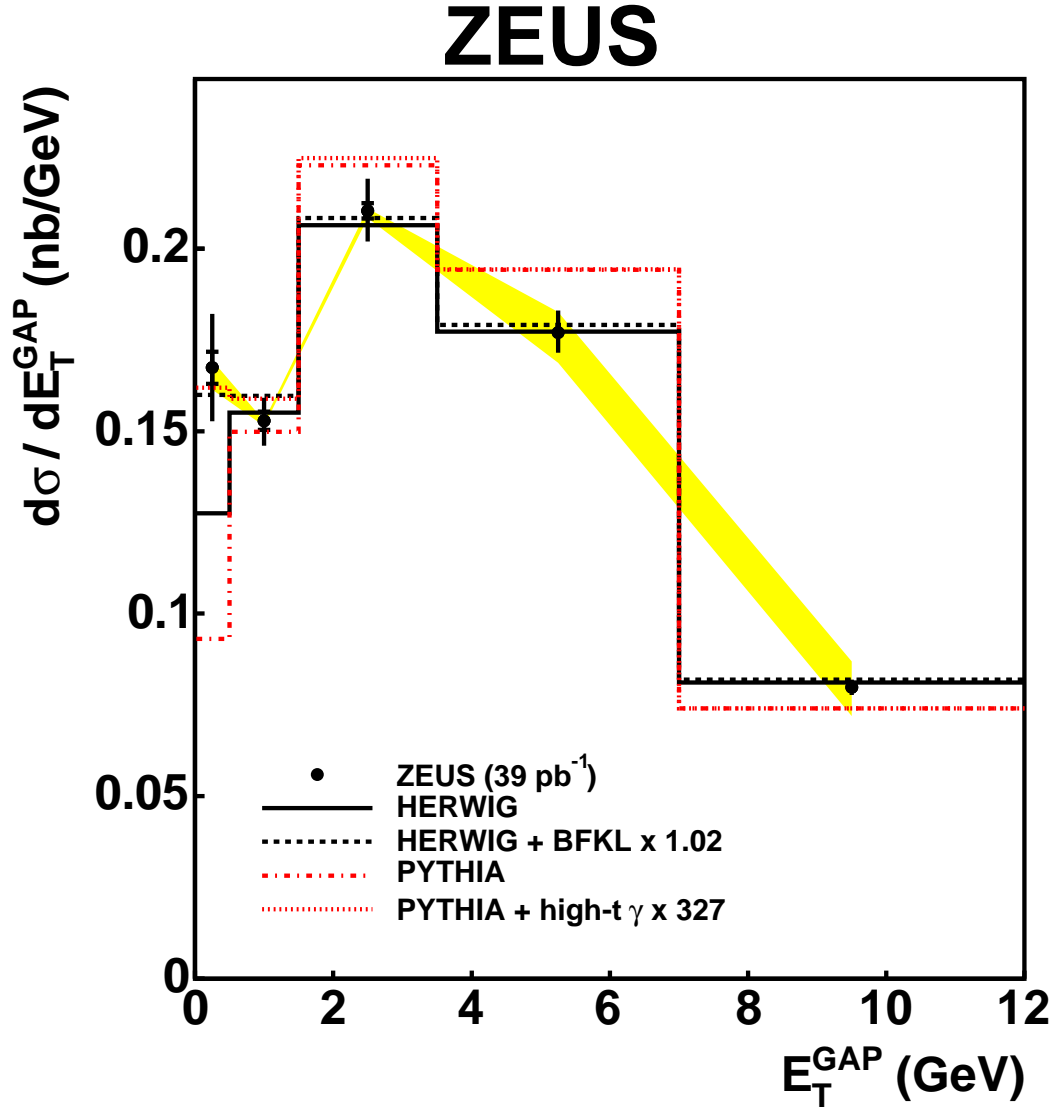,height=15cm}}
\caption{
The inclusive dijet cross section, differential in $E_{\rm T}^{\rm GAP}$.
The black circles represent the ZEUS data, with the inner error
bars representing the statistical errors and the outer error bars
representing the statistical and systematic uncertainties added in
quadrature.  The solid black line shows the prediction of
{\sc Herwig} and the black dashed line shows the prediction of
{\sc Herwig} plus {\sc BFKL} Pomeron exchange.  The dot-dashed
line shows the prediction of
{\sc Pythia} and the dotted line shows the prediction of
{\sc Pythia} plus high-$t$ photon exchange.
The band shows the calorimeter energy-scale uncertainty.
}
\label{fig:et:inc}
\end{figure}

The inclusive dijet cross section 
 as a function of $E_{\rm T}^{\rm GAP}$ 
is presented in Fig.~\ref{fig:et:inc} and Table~\ref{tbl:data:et:avg:inc}. 
At low $\etgapmath$ values, where the CS contribution should be most pronounced, 
the data demonstrate a clear excess over the NCS MC predictions.
In order to estimate the amount of CS contribution,   
the direct and resolved components of each MC were mixed according to their 
predicted MC cross
sections to give the NCS MC sample.  
The NCS and CS MC samples
were then fitted to the data according to

\[
 \frac {d\sigma} {dE_{\rm T}^{\rm GAP}} =P_1 \frac {d\sigma^{\rm NCS}} {dE_{\rm T}^{\rm GAP}} + 
  P_2 \frac {d\sigma^{\rm CS}}{dE_{\rm T}^{\rm GAP}},
\]

where $P_1$ and $P_2$ were the free parameters of the fit. The best fit to the
data resulted in $P_1=1.31 \pm 0.01 $ and $P_2=327 \pm 20$ for {\sc Pythia} and
$P_1=1.93 \pm 0.01$ and 
$P_2=1.02 \pm 0.13 $ for {\sc Herwig}.  
These scaling parameters were used in this analysis when comparing the 
data to the MC predictions.
The large value of $P_2$ for {\sc Pythia}
reflects the very low cross section of the 
high-$t$ photon exchange, which is not expected 
to represent
the mechanism of strongly-interacting CS exchange.
The color singlet contribution to the total cross section, 
estimated by integrating the MC predictions over the entire $\etgapmath$ range, 
was $(2.75 \pm 0.10)\%$ for {\sc Pythia} and $(2.04 \pm 0.25)\%$ for
{\sc Herwig}, where the errors represent only 
the statistical uncertainties of the fit.

\begin{figure}
\centerline{\epsfig{file=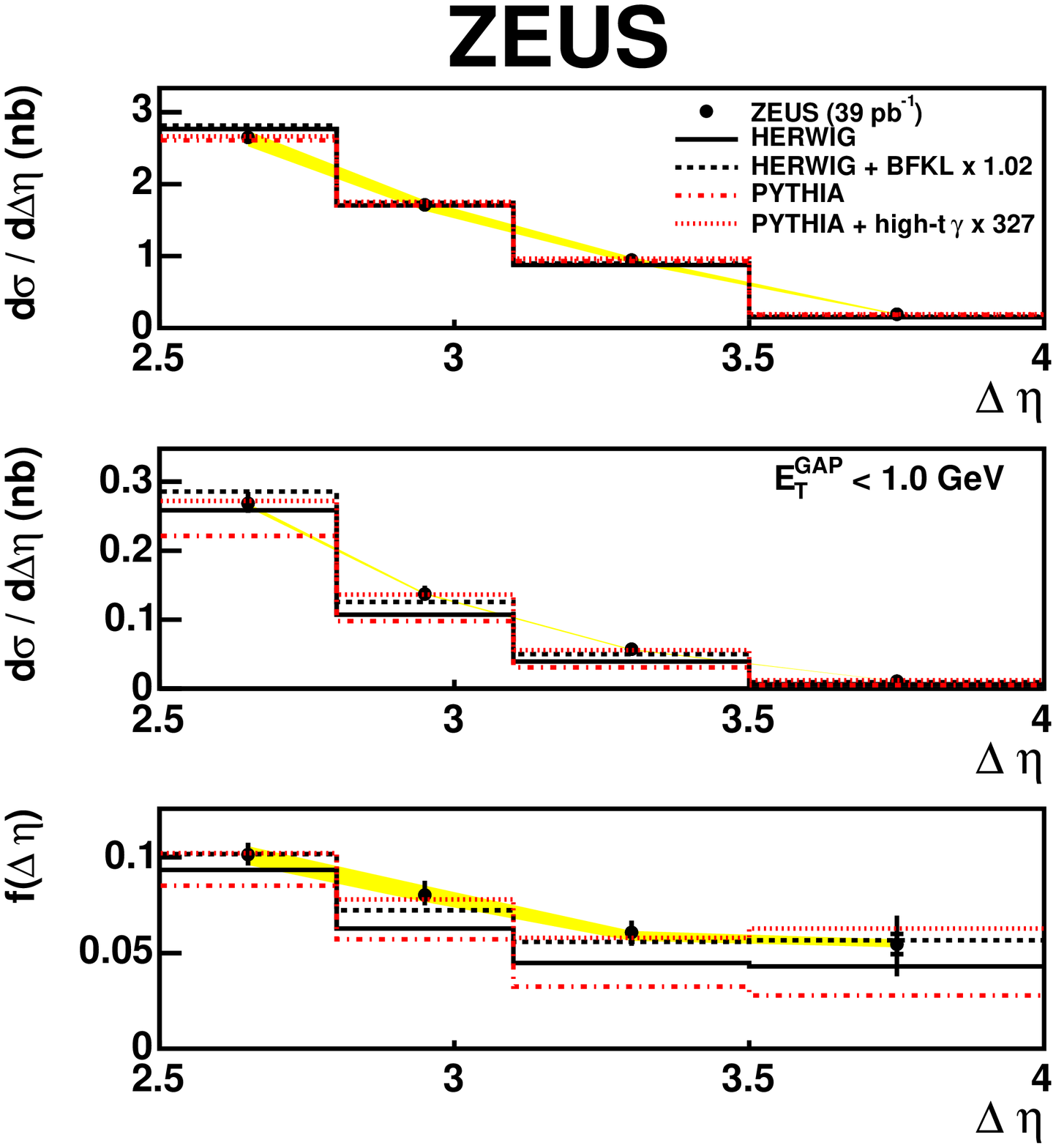,height=15cm}}
\caption{
The top plot is the inclusive dijet cross section differential in $\Delta\eta$,  
the middle plot is the gap cross section differential in $\Delta\eta$ requiring that
$E_{\rm T}^{\rm GAP} <  1 \gev$,
and the bottom plot is the gap fraction, $f$, as a function of $\Delta\eta$.
Other details as in Fig.~\protect\ref{fig:et:inc}.
}
\label{fig:de:gap}
\end{figure}

The inclusive dijet cross section, the gap cross section,
and the gap fraction as a function of the separation of the two leading jets, $\Delta \eta$,
are presented
in Fig.~\ref{fig:de:gap} 
for $E_{\rm T}^{\rm CUT}=1 {\rm \gev}$. Both cross sections and gap fractions decrease as a function of $\Delta\eta$.
In the inclusive cross section, both MC models with and without CS 
exchange describe the data equally well. For the
gap cross section the MC models without CS exchange fall below the
data, while
the MC models with CS exchange agree with the data.  
The contribution of CS exchange to the total 
gap fraction 
increases 
as the dijet separation increases from 2.5 to 4 units in pseudorapidity.

\begin{figure}
\centerline{\epsfig{file=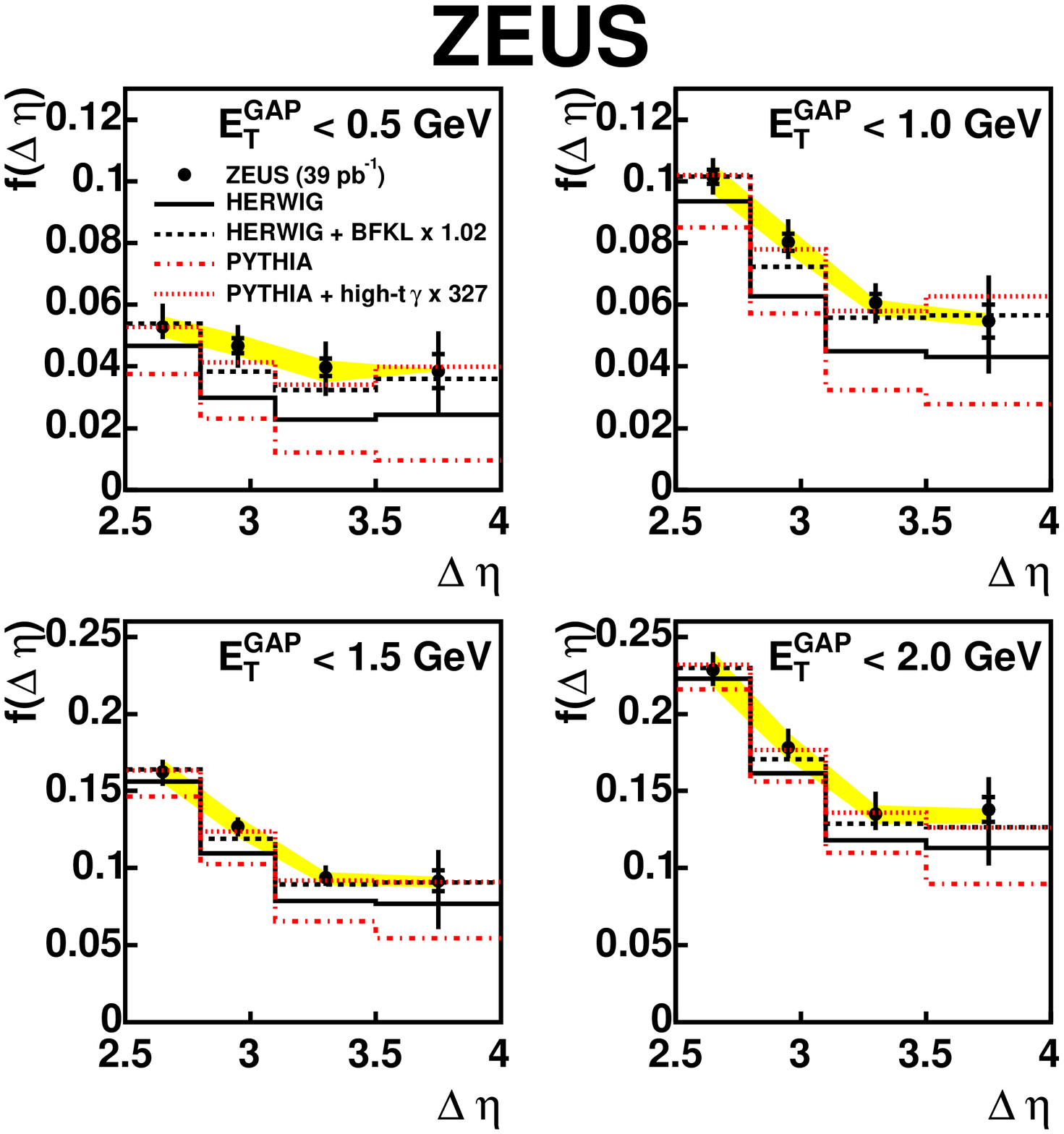,height=15cm}}
\caption{
The gap fraction, $f$, as a function of $\Delta\eta$
for different requirements on $E^{\rm GAP}_{\rm T}$.
Other details as in Fig.~\protect\ref{fig:et:inc}.
}
\label{fig:de:frac}
\end{figure}

Figure~\ref{fig:de:frac} shows the gap fraction 
as a function of $\Delta \eta$ 
for the four values of $E^{\rm CUT}_T = 0.5, 1.0, 1.5$ and $2 \gev$.   
The corresponding values are listed in Table~\ref{tbl:data:de:avg:frac}.
The data first fall and then level out as $\Delta \eta$ increases 
for all values of $E^{\rm CUT}_T$, although for
$E^{\rm CUT}_T = 0.5$ the data are consistent with a flat distribution in $\Delta \eta$.
The predictions of {\sc Pythia} and {\sc Herwig} without CS exchange lie 
below the data over the entire $\Delta \eta$ range. With the addition of 
the CS contribution, both MC models describe the data well.

\begin{figure}
\centerline{\epsfig{file=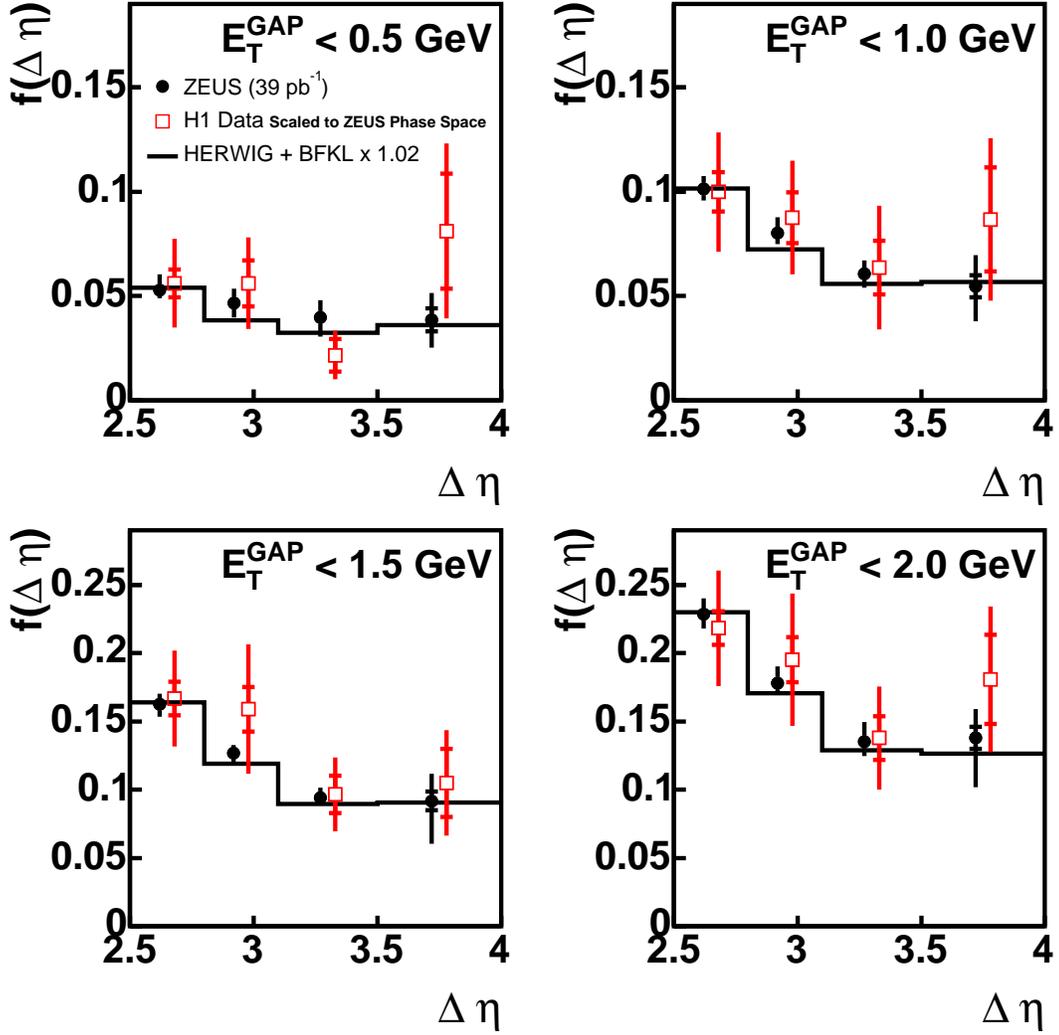,height=15cm}}
\caption{
The gap fraction, $f$, as a function of $\Delta\eta$
for different requirements on $E^{\rm GAP}_{\rm T}$.
The black circles represent the ZEUS data, with the inner error
bars representing the statistical errors and the outer error bars
representing the statistical and systematic uncertainties added in
quadrature.  The solid black line shows the prediction of
{\sc Herwig} plus {\sc BFKL} Pomeron exchange.  
The open squares represent the H1 data~\protect\cite{epj:c24:517} scaled
for comparison to the ZEUS phase space as described in the text.
}
\label{fig:de:frac:h1}
\end{figure}

The previously published ZEUS results~\cite{pl:b369:55} used a different 
definition of the rapidity gap and so cannot be directly compared.
The present results agree 
with the previous H1 measurement~\cite{epj:c24:517}, where the gap
definition used the transverse energy in the gap as for the current analysis, but with slightly 
different kinematic cuts. The comparison is shown in Fig.~\ref{fig:de:frac:h1}, where the
H1 data have been scaled bin-by-bin with multiplicative factors estimated using the 
{\sc Herwig} MC predictions for the gap fractions at the hadron level 
to account for the difference in the phase space between the ZEUS
and H1 analyses.   

\begin{figure}
\centerline{\epsfig{file=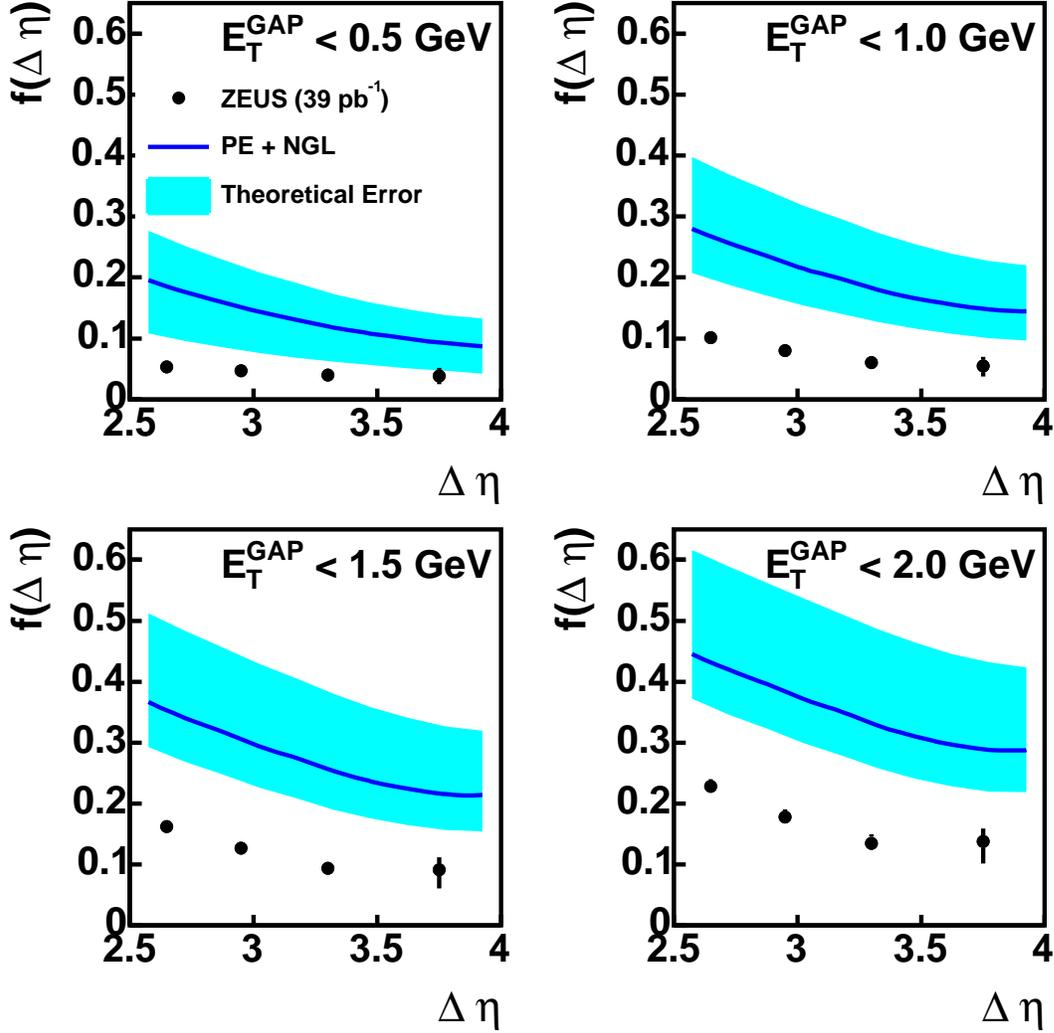,height=15cm}}
\caption{
The gap fraction, $f$, as a function of $\Delta\eta$
for different requirements on $E^{\rm GAP}_{\rm T}$.
The black circles represent the ZEUS data, with the error
bars
representing the statistical and systematic uncertainties added in
quadrature.  
The resummed calculation \protect\cite{newAppleby} 
is shown by the solid curve and
the renormalization scale uncertainty is shown by the shaded band.  The
data are plotted at the 4 bin centers in $\Delta\eta$ and the theory
curve was produced by joining the bin centers for the ratios of the integrated cross
sections for 8 bins in $\Delta\eta$.
}
\label{fig:de:frac_resum}
\end{figure}

Figure~\ref{fig:de:frac_resum} shows the gap fraction for four different values of 
$\etcutmath$ compared to the resummed calculation \cite{newAppleby}.
The shape of the data as a function of $\Delta\eta$ is reasonably well 
described for all values of $\etcutmath$
but the predictions lie above the data, almost everywhere outside of the
range defined by the theoretical uncertainties.

\begin{figure}
\centerline{\epsfig{file=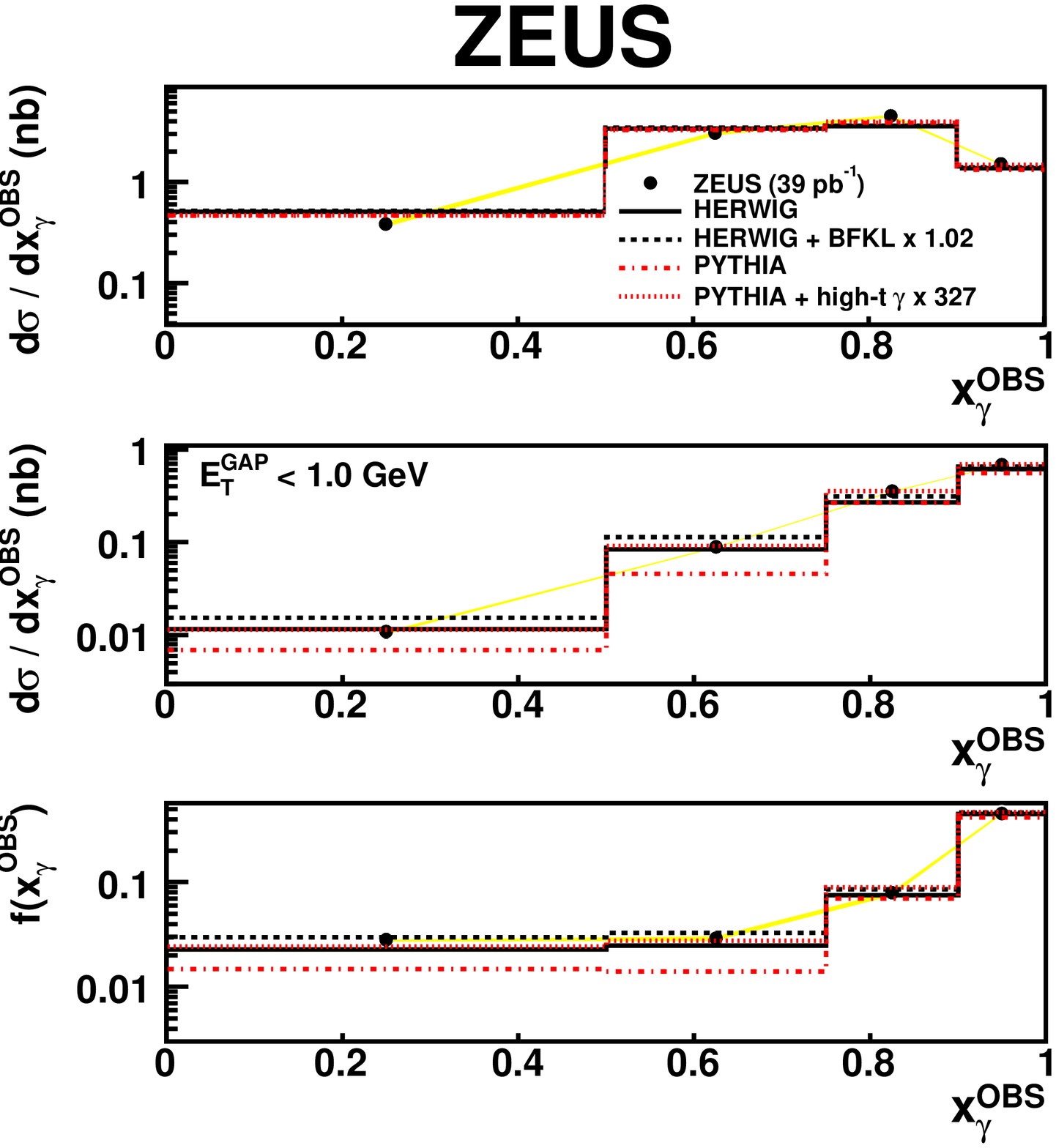,height=15cm}}
\caption{
The top plot is the inclusive dijet cross section differential in $\xgom$,  
the middle plot is the gap cross section differential in $\xgom$ requiring that
$E_{\rm T}^{\rm GAP} <  1 \gev$,
and the bottom plot is the gap fraction, $f$, as a function of $\xgom$.
Other details as in Fig.~\protect\ref{fig:et:inc}.
}
\label{fig:xg:gap}
\end{figure}

\begin{figure}
\centerline{\epsfig{file=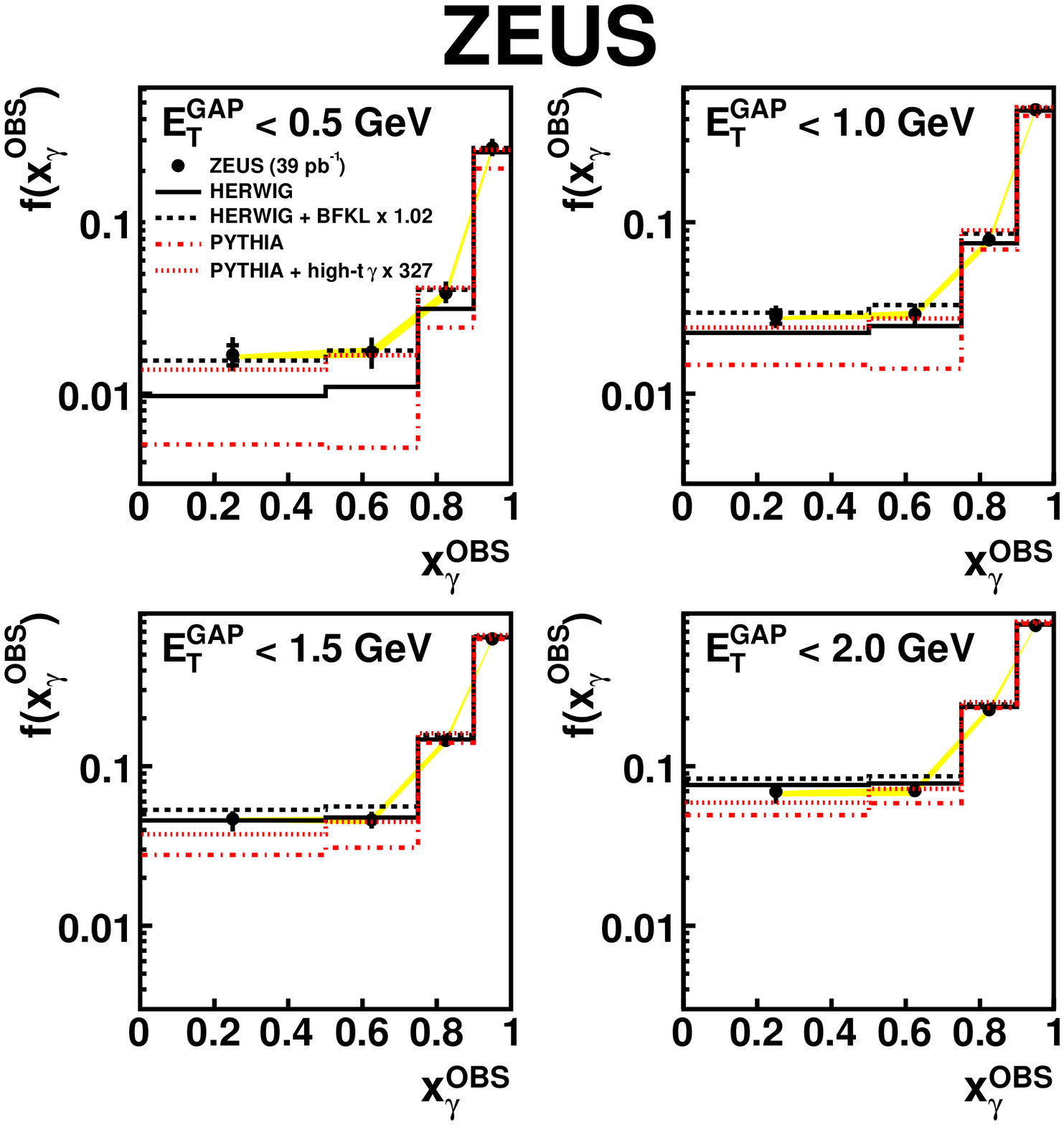,height=15cm}}
\caption{
The gap fraction, $f$, as a function of $\xgom$
for different requirements on $E^{\rm GAP}_{\rm T}$.
Other details as in Fig.~\protect\ref{fig:et:inc}.
}
\label{fig:xg:frac}
\end{figure}

For comparison with other experiments and $p \bar p$ measurements, 
which are expected to be similar to the resolved-photon 
process, the cross sections and gap fraction were
also measured as function of $\xgom$.  
These results are presented in Figs.~\ref{fig:xg:gap},~\ref{fig:xg:frac} and 
Table~\ref{tbl:data:xg:avg:frac} for four different values of $\etcutmath$.
The gap fraction decreases with decreasing $\xgom$ and 
the data are reasonably described by both MC models only
after including the CS contribution, especially
in the resolved photon region, $\xgom < 0.75$, and at low $\etgapmath$.

\begin{figure}
\centerline{\epsfig{file=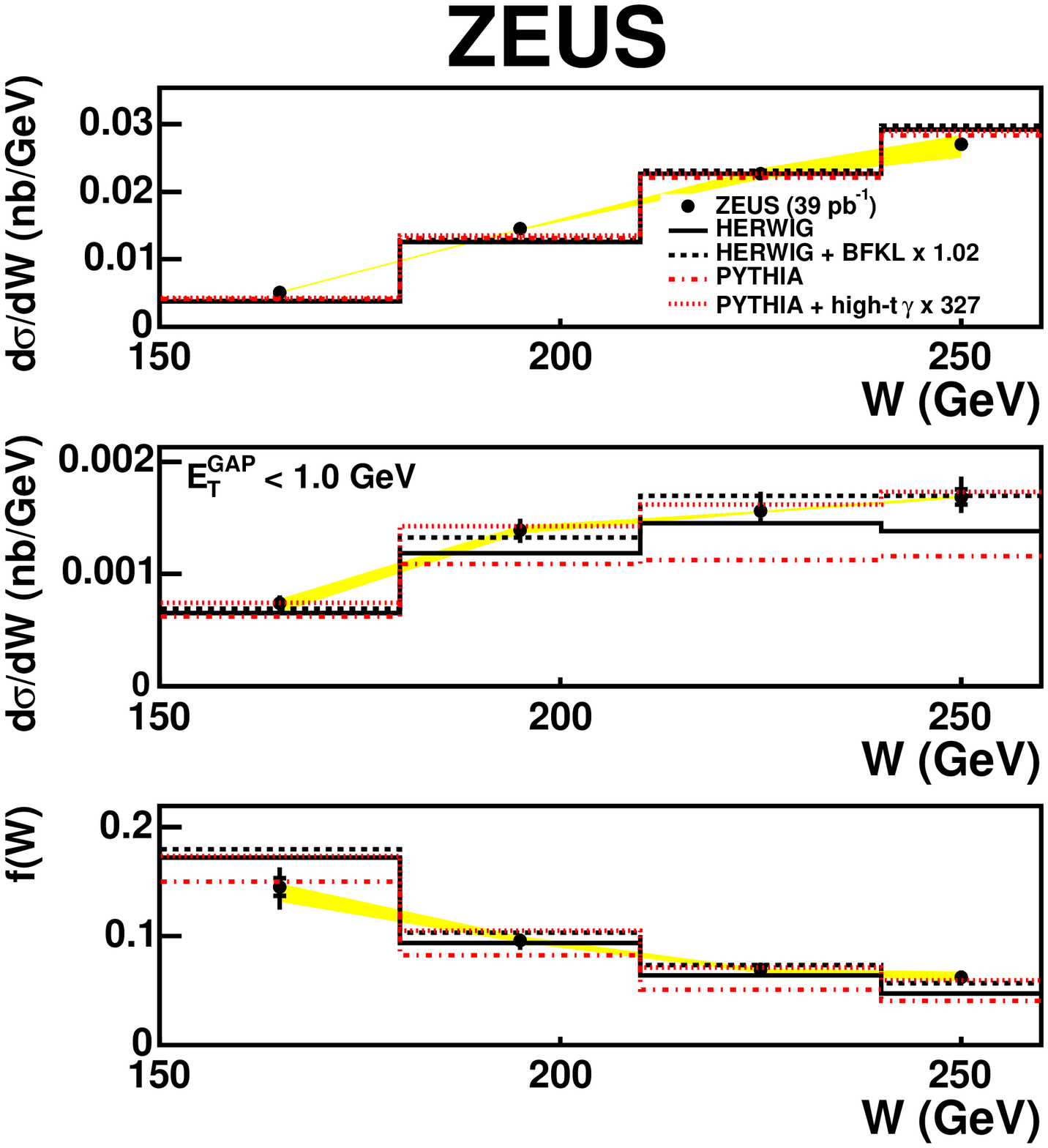,height=15cm}}
\caption{
The top plot is the inclusive dijet cross section differential in $W$,  
the middle plot is the gap cross section differential as a function of $W$ requiring that
$E_{\rm T}^{\rm GAP} <  1 \gev$,
and the bottom plot is the gap fraction, $f$, as a function of $W$.
Other details as in Fig.~\protect\ref{fig:et:inc}.
}
\label{fig:w:gap}
\end{figure}

\begin{figure}
\centerline{\epsfig{file=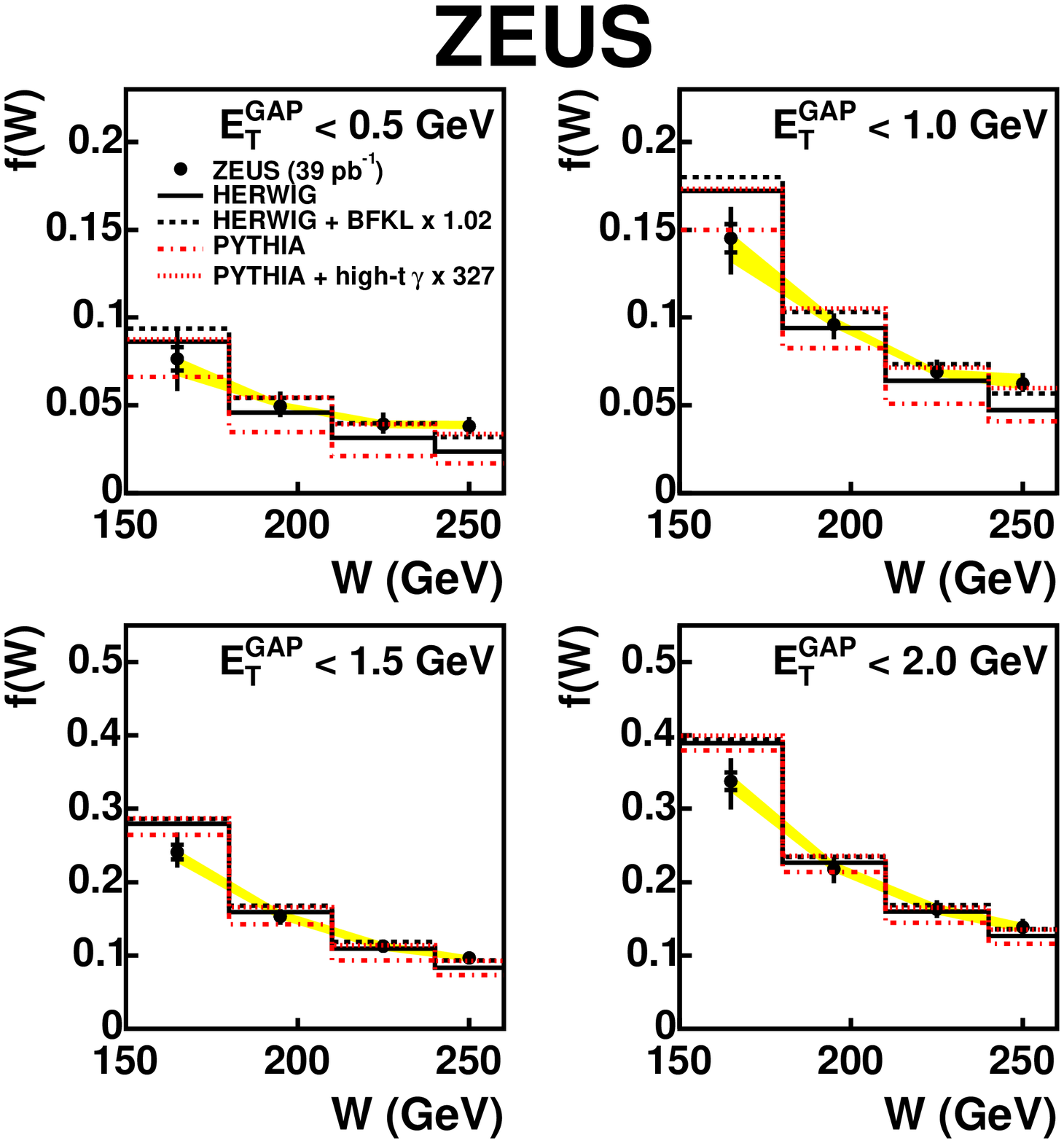,height=15cm}}
\caption{
The gap fraction, $f$, as a function of $W$
for different requirements on $E^{\rm GAP}_{\rm T}$.
Other details as in Fig.~\protect\ref{fig:et:inc}.
}
\label{fig:w:frac}
\end{figure}

The $W$ dependence, which is important for comparison with 
experiments at different energies, is presented
for the cross sections and gap fractions in 
Figs.~\ref{fig:w:gap},~\ref{fig:w:frac} and
Table~\ref{tbl:data:W:avg:frac}.  
The gap fraction falls with increasing $W$.
Both the cross sections and the gap fractions are described by the MC with CS included.

\begin{figure}
\centerline{\epsfig{file=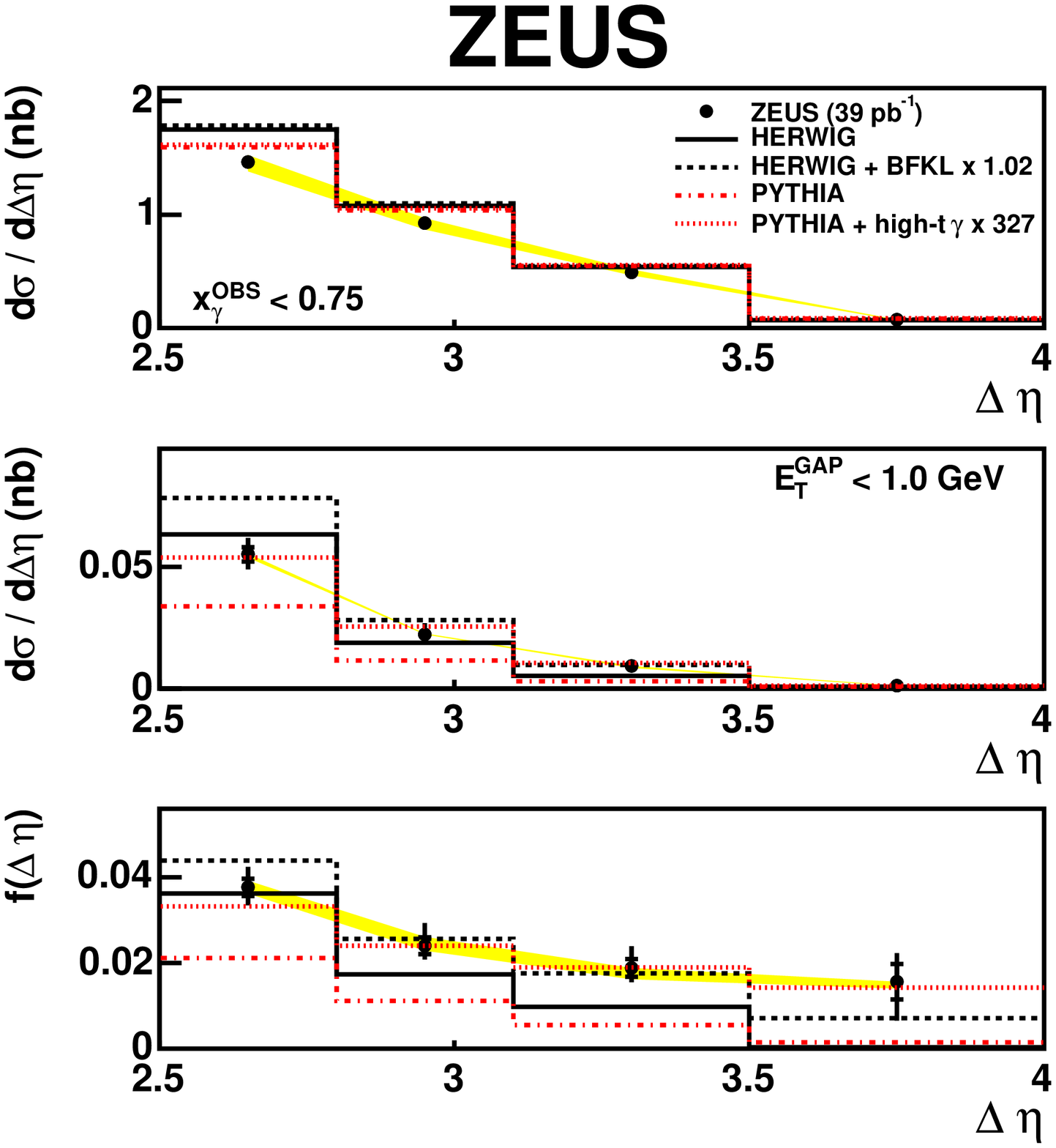,height=15cm}}
\caption{
The top plot is the inclusive dijet cross section for 
$x_{\gamma}^{\rm OBS} <0.75$ differential in $\Delta\eta$,  
the middle plot is the corresponding gap cross section differential 
in $\Delta\eta$ requiring that $E_{\rm T}^{\rm GAP} <  1 \gev$,
and the bottom plot is the gap fraction, $f$, as a function of $\Delta\eta$.
Other details as in Fig.~\protect\ref{fig:et:inc}.
}
\label{fig:de_res:gap}
\end{figure}

\begin{figure}
\centerline{\epsfig{file=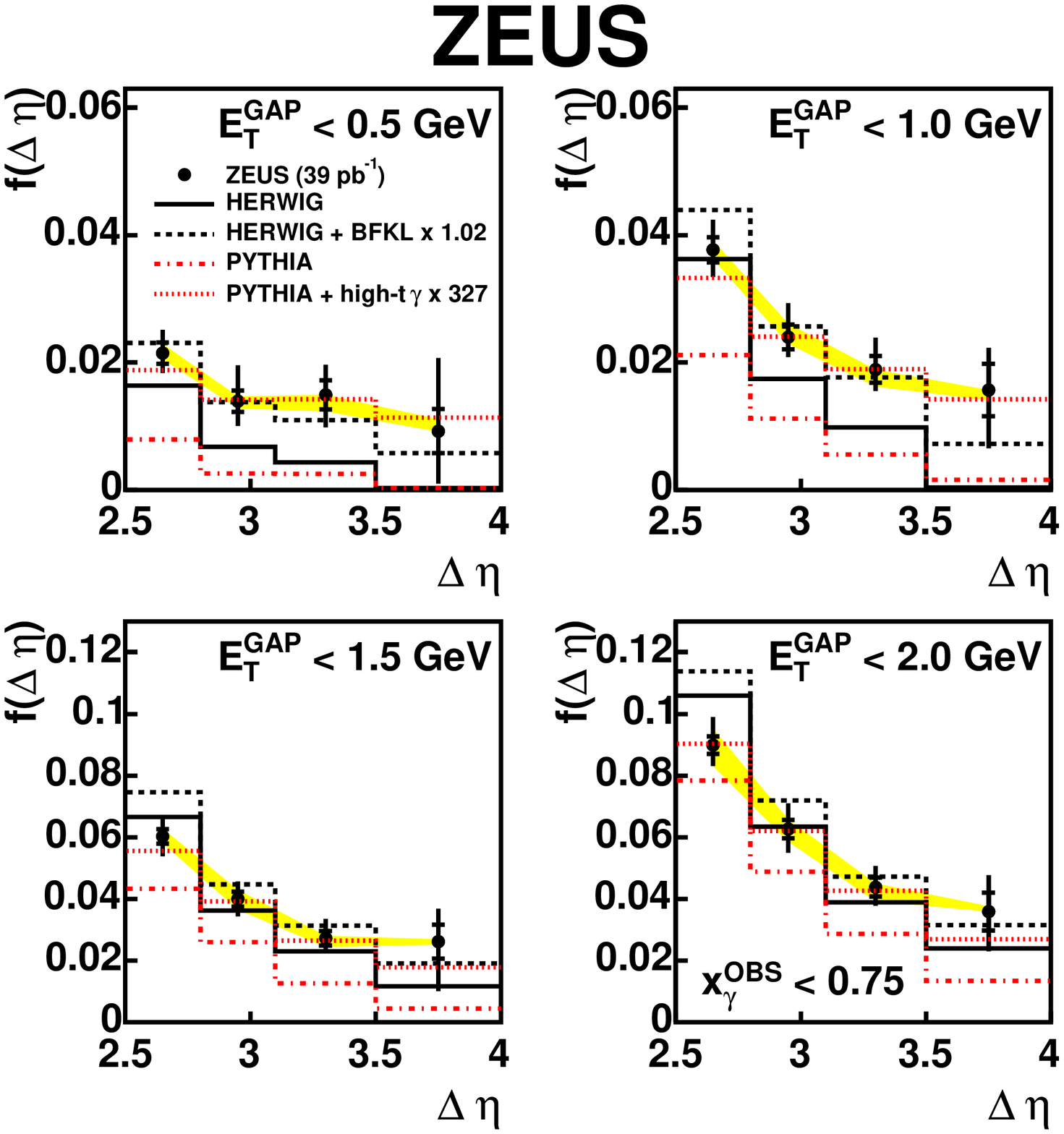,height=15cm}}
\caption{
The gap fraction, $f$, as a function of $\Delta\eta$ for $\xgom < 0.75$
and different requirements on $E^{\rm GAP}_{\rm T}$.
Other details as in Fig.~\protect\ref{fig:et:inc}.
}
\label{fig:de_res:frac}
\end{figure}

\begin{figure}
\centerline{\epsfig{file=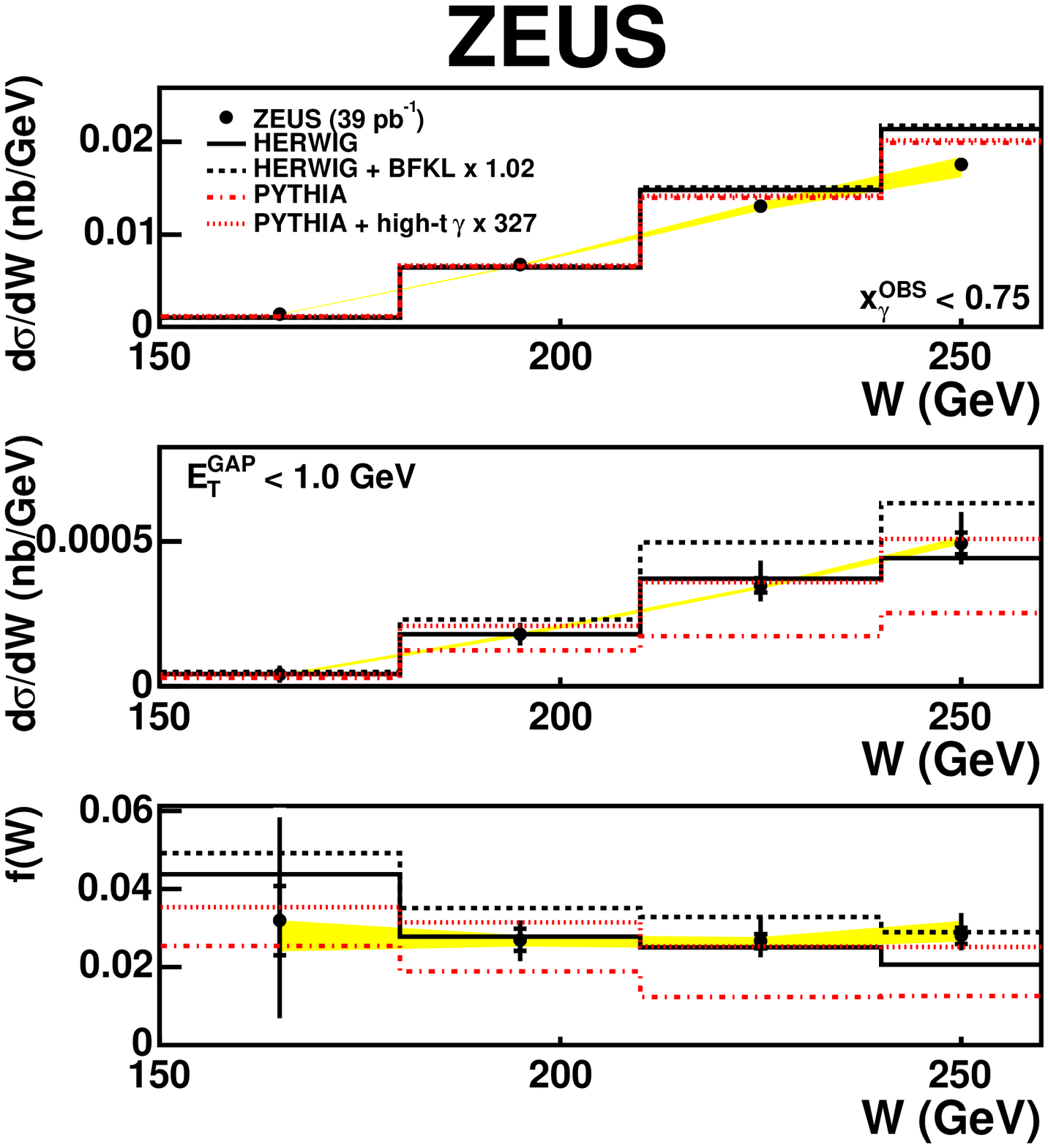,height=15cm}}
\caption{
The top plot is the inclusive dijet cross section 
for $\xgom < 0.75$ differential in $W$,  
the middle plot is the corresponding 
gap cross section differential in $W$ requiring that
$E_{\rm T}^{\rm GAP} <  1 \gev$,
and the bottom plot is the gap fraction, $f$, as a function of $W$.
Other details as in Fig.~\protect\ref{fig:et:inc}.
}
\label{fig:w_res:gap}
\end{figure}

\begin{figure}
\centerline{\epsfig{file=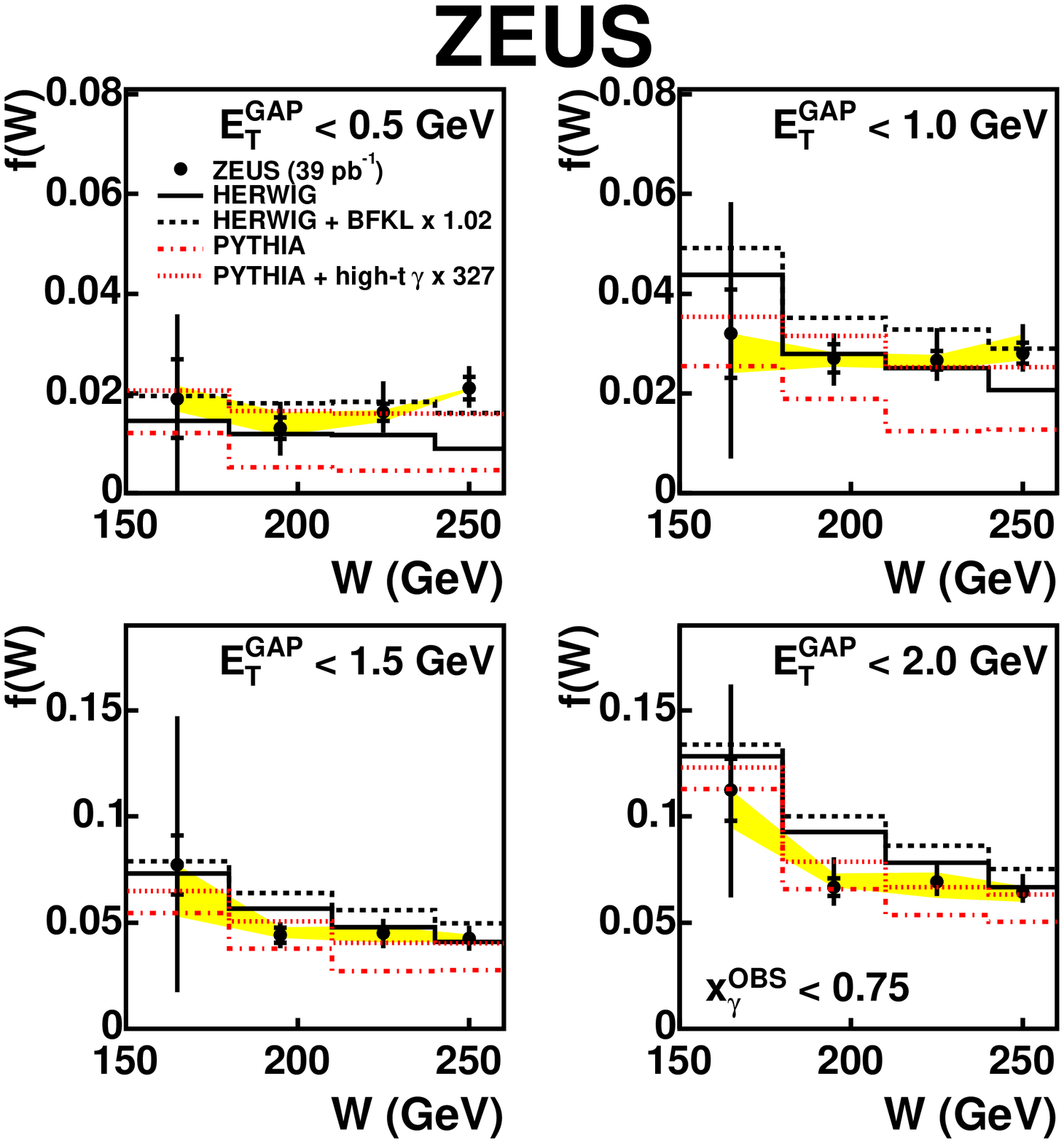,height=15cm}}
\caption{
The gap fraction, $f$, as a function of $W$, for $\xgom < 0.75$
and different requirements on $E^{\rm GAP}_{\rm T}$.
Other details as in Fig.~\protect\ref{fig:et:inc}.
}
\label{fig:w_res:frac}
\end{figure}

The  $\Delta\eta$ and $W$ 
dependencies were investigated in the 
resolved enhanced region.  Figure~\ref{fig:de_res:gap} shows the cross sections as a function of
 $\Delta\eta$ in the resolved photon region, $\xgom < 0.75$, for $\etgapmath < 1 \gev$. 
 The gap fraction as a function of $\Delta\eta$ is reasonably well described by MC
 models after including the CS contribution.
 Figure~\ref{fig:de_res:frac} and Table~\ref{tbl:data:de_res:avg:frac} show the gap 
 fractions as a function of 
$\Delta\eta$ for the resolved enhanced sample for the four $\etcutmath$ values.  
For $\etgapmath < 0.5\gev$ and $\etgapmath < 1.0\gev$,
both MC models predict almost no contribution to the gap fractions from the NCS 
component at high values of $\Delta\eta$. 
The $W$ behavior in the
resolved enhanced sample is presented in Figs.~\ref{fig:w_res:gap} and \ref{fig:w_res:frac} and 
Table~\ref{tbl:data:W_res:avg:frac}.

Although the gap fraction was measured with small errors, 
the difference in the model predictions precludes a 
model-independent determination
of the CS contribution.

\section{Summary}

Dijet photoproduction has been measured for configurations in which the 
two jets with highest transverse energy are separated by a large rapidity 
gap. The fraction of events with very little transverse energy 
between the jets is inconsistent 
with the predictions of standard photoproduction MC models. 
The same models with the inclusion of a color-singlet exchange sample at 
the level of $2 - 3\%$ 
are able to describe the data, including the gap-fraction dependency on
$\etgapmath$, $W$, $\xgom$ and $\Delta\eta$.

The difference in the model predictions precludes an accurate determination
of the color-singlet contribution and its behavior as a function of
different kinematic variables such as $\xgom$ or $W$.

\section*{Acknowledgements}
It is a pleasure to thank the DESY Directorate
for their strong support and encouragement. The remarkable achievements of the
HERA machine group were essential for the successful completion of this work and
are greatly appreciated. The design, construction and installation
of the ZEUS detector has been made possible by the efforts of many people
who are not listed as authors. 
We are indebted to R. Appleby and M. Dasgupta for invaluable 
discussions and for providing the resummed calculation.

\vfill\eject

{
\def\bibname{\Large\bf References}
\def\refname{\Large\bf References}
\pagestyle{plain}
\ifzeusbst
  \bibliographystyle{./BiBTeX/bst/l4z_default}
\fi
\ifzdrftbst
  \bibliographystyle{./BiBTeX/bst/l4z_draft}
\fi
\ifzbstepj
  \bibliographystyle{./BiBTeX/bst/l4z_epj}
\fi
\ifzbstnp
  \bibliographystyle{./BiBTeX/bst/l4z_np}
\fi
\ifzbstpl
  \bibliographystyle{./BiBTeX/bst/l4z_pl}
\fi
{\raggedright
\bibliography{./BiBTeX/user/syn.bib,%
              ./BiBTeX/bib/l4z_articles.bib,%
              ./BiBTeX/bib/l4z_books.bib,%
              ./BiBTeX/bib/l4z_conferences.bib,%
              ./BiBTeX/bib/l4z_h1.bib,%
              ./BiBTeX/bib/l4z_misc.bib,%
              ./BiBTeX/bib/l4z_old.bib,%
              ./BiBTeX/bib/l4z_preprints.bib,%
              ./BiBTeX/bib/l4z_replaced.bib,%
              ./BiBTeX/bib/l4z_temporary.bib,%
              ./BiBTeX/bib/l4z_zeus.bib}}
}
\vfill\eject

\begin{table}
\begin{center}
\begin{tabular}{|c|cccc|} 
\hline 
$E_{\rm T}^{\rm GAP}$ bin $(\gev)$ &  $\sigma (\nb/\gev)$ &  $\pm$ stat &  $\pm$ sys &  $\pm$ cal \\ 
\hline \hline 
$   0.0-0.5$ & 0.167 & $\pm$ 0.004 & $^{+ 0.014}_{- 0.014}$  & $^{+ 0.002}_{- 0.006}$ \\
$   0.5-1.5$ & 0.153 & $\pm$ 0.002 & $^{+ 0.006}_{- 0.006}$  & $^{+ 0.000}_{- 0.001}$ \\
$   1.5-3.5$ & 0.210 & $\pm$ 0.002 & $^{+ 0.009}_{- 0.008}$  & $^{+ 0.001}_{- 0.002}$ \\
$   3.5-7.0$ & 0.177 & $\pm$ 0.001 & $^{+ 0.006}_{- 0.005}$  & $^{+ 0.006}_{- 0.008}$ \\
$   7.0-12.0$ & 0.080 & $\pm$ 0.001 & $^{+ 0.002}_{- 0.002}$  & $^{+ 0.007}_{- 0.008}$ \\
\hline
\end{tabular}
\caption{The measured differential cross section $d\sigma / d E_{\rm T}^{\rm GAP}$
 unfolded with the average correction factors of {\sc Pythia} and {\sc Herwig} for the inclusive sample of events.
The statistical error, systematic errors, and calorimeter energy-scale uncertainty  on the measurement are also listed.}
\label{tbl:data:et:avg:inc}
\end{center}
\end{table}
\begin{table}
\begin{center}
\begin{tabular}{|c|c|cccc|} 
\hline 
$\Delta \eta$ bin & $E_{\rm T}^{\rm CUT} \gev$ &  $f$ &  $\pm$ stat &  $\pm$ sys &  $\pm$ cal \\ 
\hline \hline 
$    2.5,2.8$ & \multirow{4}*{0.5} & 0.053 & $\pm$ 0.002 & $^{+ 0.007}_{- 0.004}$  & $^{+ 0.003}_{- 0.003}$ \\
$    2.8,3.1$ &  & 0.047 & $\pm$ 0.002 & $^{+ 0.006}_{- 0.007}$  & $^{+ 0.004}_{- 0.003}$ \\
$    3.1,3.5$ &  & 0.040 & $\pm$ 0.003 & $^{+ 0.008}_{- 0.009}$  & $^{+ 0.002}_{- 0.005}$ \\
$    3.5,4.0$ &  & 0.038 & $\pm$ 0.005 & $^{+ 0.012}_{- 0.012}$  & $^{+ 0.001}_{- 0.000}$ \\
\hline
$    2.5,2.8$ & \multirow{4}*{1.0} & 0.101 & $\pm$ 0.002 & $^{+ 0.006}_{- 0.005}$  & $^{+ 0.004}_{- 0.005}$ \\
$    2.8,3.1$ &  & 0.080 & $\pm$ 0.003 & $^{+ 0.007}_{- 0.005}$  & $^{+ 0.005}_{- 0.004}$ \\
$    3.1,3.5$ &  & 0.061 & $\pm$ 0.003 & $^{+ 0.006}_{- 0.006}$  & $^{+ 0.001}_{- 0.004}$ \\
$    3.5,4.0$ &  & 0.055 & $\pm$ 0.005 & $^{+ 0.014}_{- 0.016}$  & $^{+ 0.003}_{- 0.002}$ \\
\hline
$    2.5,2.8$ & \multirow{4}*{1.5} & 0.163 & $\pm$ 0.003 & $^{+ 0.007}_{- 0.009}$  & $^{+ 0.008}_{- 0.007}$ \\
$    2.8,3.1$ &  & 0.127 & $\pm$ 0.003 & $^{+ 0.005}_{- 0.005}$  & $^{+ 0.007}_{- 0.007}$ \\
$    3.1,3.5$ &  & 0.094 & $\pm$ 0.003 & $^{+ 0.007}_{- 0.005}$  & $^{+ 0.003}_{- 0.005}$ \\
$    3.5,4.0$ &  & 0.092 & $\pm$ 0.007 & $^{+ 0.019}_{- 0.030}$  & $^{+ 0.003}_{- 0.004}$ \\
\hline
$    2.5,2.8$ & \multirow{4}*{2.0} & 0.228 & $\pm$ 0.003 & $^{+ 0.011}_{- 0.010}$  & $^{+ 0.012}_{- 0.011}$ \\
$    2.8,3.1$ &  & 0.178 & $\pm$ 0.004 & $^{+ 0.012}_{- 0.006}$  & $^{+ 0.010}_{- 0.008}$ \\
$    3.1,3.5$ &  & 0.135 & $\pm$ 0.004 & $^{+ 0.014}_{- 0.010}$  & $^{+ 0.006}_{- 0.006}$ \\
$    3.5,4.0$ &  & 0.138 & $\pm$ 0.008 & $^{+ 0.019}_{- 0.035}$  & $^{+ 0.001}_{- 0.009}$ \\
\hline
\end{tabular}
\caption{The measured gap fraction $f\left( \Delta \eta \right)$
 unfolded with the average correction factors of {\sc Pythia} and {\sc Herwig}.
The statistical error, systematic errors, and calorimeter energy-scale uncertainty  on the measurement are also listed.}
\label{tbl:data:de:avg:frac}
\end{center}
\end{table}
\begin{table}
\begin{center}
\begin{tabular}{|c|c|cccc|} 
\hline 
$x_{\gamma}^{\rm OBS}$ bin & $E_{\rm T}^{\rm CUT} \gev$ &  $f$ &  $\pm$ stat &  $\pm$ sys &  $\pm$ cal \\ 
\hline \hline 
$   0.00,0.50$ & \multirow{4}*{0.5} & 0.017 & $\pm$ 0.002 & $^{+ 0.004}_{- 0.002}$  & $^{+ 0.000}_{- 0.001}$ \\
$   0.50,0.75$ &  & 0.018 & $\pm$ 0.001 & $^{+ 0.004}_{- 0.003}$  & $^{+ 0.001}_{- 0.001}$ \\
$   0.75,0.90$ &  & 0.039 & $\pm$ 0.002 & $^{+ 0.006}_{- 0.005}$  & $^{+ 0.002}_{- 0.003}$ \\
$   0.90,1.00$ &  & 0.272 & $\pm$ 0.010 & $^{+ 0.033}_{- 0.028}$  & $^{+ 0.011}_{- 0.012}$ \\
\hline
$   0.00,0.50$ & \multirow{4}*{1.0} & 0.028 & $\pm$ 0.003 & $^{+ 0.004}_{- 0.003}$  & $^{+ 0.000}_{- 0.001}$ \\
$   0.50,0.75$ &  & 0.029 & $\pm$ 0.001 & $^{+ 0.004}_{- 0.003}$  & $^{+ 0.001}_{- 0.002}$ \\
$   0.75,0.90$ &  & 0.079 & $\pm$ 0.002 & $^{+ 0.005}_{- 0.005}$  & $^{+ 0.003}_{- 0.005}$ \\
$   0.90,1.00$ &  & 0.454 & $\pm$ 0.012 & $^{+ 0.024}_{- 0.026}$  & $^{+ 0.008}_{- 0.008}$ \\
\hline
$   0.00,0.50$ & \multirow{4}*{1.5} & 0.047 & $\pm$ 0.003 & $^{+ 0.005}_{- 0.007}$  & $^{+ 0.001}_{- 0.002}$ \\
$   0.50,0.75$ &  & 0.046 & $\pm$ 0.001 & $^{+ 0.006}_{- 0.005}$  & $^{+ 0.003}_{- 0.003}$ \\
$   0.75,0.90$ &  & 0.145 & $\pm$ 0.003 & $^{+ 0.007}_{- 0.010}$  & $^{+ 0.006}_{- 0.008}$ \\
$   0.90,1.00$ &  & 0.630 & $\pm$ 0.015 & $^{+ 0.028}_{- 0.022}$  & $^{+ 0.010}_{- 0.007}$ \\
\hline
$   0.00,0.50$ & \multirow{4}*{2.0} & 0.069 & $\pm$ 0.004 & $^{+ 0.007}_{- 0.010}$  & $^{+ 0.001}_{- 0.005}$ \\
$   0.50,0.75$ &  & 0.070 & $\pm$ 0.002 & $^{+ 0.008}_{- 0.005}$  & $^{+ 0.004}_{- 0.005}$ \\
$   0.75,0.90$ &  & 0.227 & $\pm$ 0.004 & $^{+ 0.016}_{- 0.013}$  & $^{+ 0.010}_{- 0.009}$ \\
$   0.90,1.00$ &  & 0.763 & $\pm$ 0.018 & $^{+ 0.023}_{- 0.021}$  & $^{+ 0.009}_{- 0.003}$ \\
\hline
\end{tabular}
\caption{The measured gap fraction f$\left( x_{\gamma}^{\rm OBS} \right)$
 unfolded with the average correction factors of {\sc Pythia} and {\sc Herwig}.
The statistical error, systematic errors, and calorimeter energy-scale uncertainty  on the measurement are also listed.}
\label{tbl:data:xg:avg:frac}
\end{center}
\end{table}
\begin{table}
\begin{center}
\begin{tabular}{|c|c|cccc|} 
\hline 
$W$ bin $(\gev)$ & $E_{\rm T}^{\rm CUT} \gev$ &  $f$ &  $\pm$ stat &  $\pm$ sys &  $\pm$ cal \\ 
\hline \hline 
$  150.0,180.0$ & \multirow{4}*{0.5} & 0.077 & $\pm$ 0.007 & $^{+ 0.017}_{- 0.017}$  & $^{+ 0.001}_{- 0.010}$ \\
$  180.0,210.0$ &  & 0.049 & $\pm$ 0.003 & $^{+ 0.008}_{- 0.005}$  & $^{+ 0.002}_{- 0.001}$ \\
$  210.0,240.0$ &  & 0.039 & $\pm$ 0.002 & $^{+ 0.006}_{- 0.005}$  & $^{+ 0.002}_{- 0.002}$ \\
$  240.0,260.0$ &  & 0.038 & $\pm$ 0.002 & $^{+ 0.005}_{- 0.004}$  & $^{+ 0.003}_{- 0.002}$ \\
\hline
$  150.0,180.0$ & \multirow{4}*{1.0} & 0.145 & $\pm$ 0.008 & $^{+ 0.016}_{- 0.019}$  & $^{+ 0.003}_{- 0.014}$ \\
$  180.0,210.0$ &  & 0.096 & $\pm$ 0.004 & $^{+ 0.005}_{- 0.007}$  & $^{+ 0.004}_{- 0.001}$ \\
$  210.0,240.0$ &  & 0.069 & $\pm$ 0.002 & $^{+ 0.007}_{- 0.004}$  & $^{+ 0.001}_{- 0.002}$ \\
$  240.0,260.0$ &  & 0.062 & $\pm$ 0.002 & $^{+ 0.006}_{- 0.004}$  & $^{+ 0.005}_{- 0.003}$ \\
\hline
$  150.0,180.0$ & \multirow{4}*{1.5} & 0.241 & $\pm$ 0.010 & $^{+ 0.025}_{- 0.019}$  & $^{+ 0.003}_{- 0.015}$ \\
$  180.0,210.0$ &  & 0.153 & $\pm$ 0.004 & $^{+ 0.010}_{- 0.010}$  & $^{+ 0.008}_{- 0.004}$ \\
$  210.0,240.0$ &  & 0.113 & $\pm$ 0.003 & $^{+ 0.008}_{- 0.008}$  & $^{+ 0.006}_{- 0.006}$ \\
$  240.0,260.0$ &  & 0.097 & $\pm$ 0.003 & $^{+ 0.006}_{- 0.006}$  & $^{+ 0.003}_{- 0.005}$ \\
\hline
$  150.0,180.0$ & \multirow{4}*{2.0} & 0.338 & $\pm$ 0.012 & $^{+ 0.029}_{- 0.037}$  & $^{+ 0.010}_{- 0.015}$ \\
$  180.0,210.0$ &  & 0.218 & $\pm$ 0.005 & $^{+ 0.016}_{- 0.019}$  & $^{+ 0.010}_{- 0.004}$ \\
$  210.0,240.0$ &  & 0.163 & $\pm$ 0.003 & $^{+ 0.012}_{- 0.011}$  & $^{+ 0.007}_{- 0.007}$ \\
$  240.0,260.0$ &  & 0.139 & $\pm$ 0.003 & $^{+ 0.011}_{- 0.004}$  & $^{+ 0.006}_{- 0.008}$ \\
\hline
\end{tabular}
\caption{The measured gap fraction $f \left( W \right)$
 unfolded with the average correction factors of {\sc Pythia} and {\sc Herwig}.
The statistical error, systematic errors, and calorimeter energy-scale uncertainty  on the measurement are also listed.}
\label{tbl:data:W:avg:frac}
\end{center}
\end{table}
\begin{table}
\begin{center}
\begin{tabular}{|c|c|cccc|} 
\hline 
$\Delta \eta$ bin & $E_{\rm T}^{\rm CUT} \gev$ &  $f$ &  $\pm$ stat &  $\pm$ sys &  $\pm$ cal \\ 
\hline \hline 
$    2.5,2.8$ & \multirow{4}*{0.5} & 0.021 & $\pm$ 0.002 & $^{+ 0.003}_{- 0.003}$  & $^{+ 0.001}_{- 0.001}$ \\
$    2.8,3.1$ &  & 0.014 & $\pm$ 0.002 & $^{+ 0.005}_{- 0.004}$  & $^{+ 0.001}_{- 0.001}$ \\
$    3.1,3.5$ &  & 0.015 & $\pm$ 0.002 & $^{+ 0.004}_{- 0.005}$  & $^{+ 0.000}_{- 0.003}$ \\
$    3.5,4.0$ &  & 0.009 & $\pm$ 0.003 & $^{+ 0.011}_{- 0.007}$  & $^{+ 0.002}_{- 0.000}$ \\
\hline
$    2.5,2.8$ & \multirow{4}*{1.0} & 0.038 & $\pm$ 0.002 & $^{+ 0.004}_{- 0.004}$  & $^{+ 0.001}_{- 0.001}$ \\
$    2.8,3.1$ &  & 0.024 & $\pm$ 0.002 & $^{+ 0.005}_{- 0.003}$  & $^{+ 0.002}_{- 0.001}$ \\
$    3.1,3.5$ &  & 0.019 & $\pm$ 0.002 & $^{+ 0.005}_{- 0.003}$  & $^{+ 0.000}_{- 0.003}$ \\
$    3.5,4.0$ &  & 0.016 & $\pm$ 0.004 & $^{+ 0.005}_{- 0.008}$  & $^{+ 0.000}_{- 0.002}$ \\
\hline
$    2.5,2.8$ & \multirow{4}*{1.5} & 0.060 & $\pm$ 0.002 & $^{+ 0.006}_{- 0.006}$  & $^{+ 0.003}_{- 0.002}$ \\
$    2.8,3.1$ &  & 0.040 & $\pm$ 0.002 & $^{+ 0.005}_{- 0.005}$  & $^{+ 0.003}_{- 0.003}$ \\
$    3.1,3.5$ &  & 0.027 & $\pm$ 0.002 & $^{+ 0.006}_{- 0.003}$  & $^{+ 0.001}_{- 0.003}$ \\
$    3.5,4.0$ &  & 0.026 & $\pm$ 0.006 & $^{+ 0.009}_{- 0.015}$  & $^{+ 0.001}_{- 0.001}$ \\
\hline
$    2.5,2.8$ & \multirow{4}*{2.0} & 0.090 & $\pm$ 0.003 & $^{+ 0.009}_{- 0.006}$  & $^{+ 0.005}_{- 0.007}$ \\
$    2.8,3.1$ &  & 0.063 & $\pm$ 0.003 & $^{+ 0.008}_{- 0.007}$  & $^{+ 0.003}_{- 0.004}$ \\
$    3.1,3.5$ &  & 0.044 & $\pm$ 0.003 & $^{+ 0.006}_{- 0.005}$  & $^{+ 0.001}_{- 0.005}$ \\
$    3.5,4.0$ &  & 0.036 & $\pm$ 0.006 & $^{+ 0.010}_{- 0.011}$  & $^{+ 0.002}_{- 0.000}$ \\
\hline
\end{tabular}
\caption{The measured gap fraction $f\left( \Delta \eta \right)$
for the region $\xgom < 0.75$
 unfolded with the average correction factors of {\sc Pythia} and {\sc Herwig}.
The statistical error, systematic errors, and calorimeter energy-scale uncertainty  on the measurement are also listed.}
\label{tbl:data:de_res:avg:frac}
\end{center}
\end{table}
\begin{table}
\begin{center}
\begin{tabular}{|c|c|cccc|} 
\hline 
$W$ bin $(\gev)$ & $E_{\rm T}^{\rm CUT} \gev$ &  $f$ &  $\pm$ stat &  $\pm$ sys &  $\pm$ cal \\ 
\hline \hline 
$  150.0,180.0$ & \multirow{4}*{0.5} & 0.019 & $\pm$ 0.008 & $^{+ 0.015}_{- 0.018}$  & $^{+ 0.003}_{- 0.003}$ \\
$  180.0,210.0$ &  & 0.013 & $\pm$ 0.002 & $^{+ 0.004}_{- 0.005}$  & $^{+ 0.003}_{- 0.001}$ \\
$  210.0,240.0$ &  & 0.016 & $\pm$ 0.002 & $^{+ 0.006}_{- 0.004}$  & $^{+ 0.000}_{- 0.002}$ \\
$  240.0,260.0$ &  & 0.021 & $\pm$ 0.002 & $^{+ 0.004}_{- 0.003}$  & $^{+ 0.000}_{- 0.001}$ \\
\hline
$  150.0,180.0$ & \multirow{4}*{1.0} & 0.032 & $\pm$ 0.009 & $^{+ 0.025}_{- 0.023}$  & $^{+ 0.000}_{- 0.008}$ \\
$  180.0,210.0$ &  & 0.027 & $\pm$ 0.003 & $^{+ 0.004}_{- 0.005}$  & $^{+ 0.001}_{- 0.002}$ \\
$  210.0,240.0$ &  & 0.027 & $\pm$ 0.002 & $^{+ 0.006}_{- 0.004}$  & $^{+ 0.001}_{- 0.002}$ \\
$  240.0,260.0$ &  & 0.028 & $\pm$ 0.002 & $^{+ 0.005}_{- 0.003}$  & $^{+ 0.004}_{- 0.001}$ \\
\hline
$  150.0,180.0$ & \multirow{4}*{1.5} & 0.077 & $\pm$ 0.014 & $^{+ 0.068}_{- 0.058}$  & $^{+ 0.000}_{- 0.024}$ \\
$  180.0,210.0$ &  & 0.044 & $\pm$ 0.004 & $^{+ 0.005}_{- 0.005}$  & $^{+ 0.004}_{- 0.001}$ \\
$  210.0,240.0$ &  & 0.045 & $\pm$ 0.002 & $^{+ 0.006}_{- 0.007}$  & $^{+ 0.003}_{- 0.004}$ \\
$  240.0,260.0$ &  & 0.043 & $\pm$ 0.002 & $^{+ 0.005}_{- 0.005}$  & $^{+ 0.002}_{- 0.002}$ \\
\hline
$  150.0,180.0$ & \multirow{4}*{2.0} & 0.113 & $\pm$ 0.015 & $^{+ 0.048}_{- 0.048}$  & $^{+ 0.000}_{- 0.018}$ \\
$  180.0,210.0$ &  & 0.067 & $\pm$ 0.004 & $^{+ 0.013}_{- 0.007}$  & $^{+ 0.006}_{- 0.000}$ \\
$  210.0,240.0$ &  & 0.069 & $\pm$ 0.003 & $^{+ 0.007}_{- 0.006}$  & $^{+ 0.004}_{- 0.008}$ \\
$  240.0,260.0$ &  & 0.064 & $\pm$ 0.003 & $^{+ 0.008}_{- 0.004}$  & $^{+ 0.003}_{- 0.005}$ \\
\hline
\end{tabular}
\caption{The measured gap fraction $f \left( W \right)$ 
for the region $\xgom < 0.75$
 unfolded with the average correction factors of {\sc Pythia} and {\sc Herwig}.
The statistical error, systematic errors, and calorimeter energy-scale uncertainty  on the measurement are also listed.}
\label{tbl:data:W_res:avg:frac}
\end{center}
\end{table}
\vfill\eject

\vfill\eject

%
\end{document}